\documentclass[pre,twocolumn,showpacs,preprintnumbers,floatfix]{revtex4}
\usepackage{graphicx,citesort,psfrag,amsmath,amssymb}

\begin{document}

\bibliographystyle{apsrev}

 
\title{Minimal model for aeolian sand dunes}

\author{Klaus Kroy}
\affiliation{Dept.\ of Physics and Astronomy, University of
Edinburgh, EH9 3JZ, United Kingdom}
\author{Gerd Sauermann, and Hans J. Herrmann}
\affiliation{ICA-1, Universit\"at Stuttgart, Pfaffenwaldring 27, 70569
Stuttgart, Germany} 

\date{\today}

\begin{abstract}
 We present a minimal model for the formation and migration of aeolian
  sand dunes. It combines a perturbative description of the turbulent
  wind velocity field above the dune with a continuum saltation model
  that allows for saturation transients in the sand flux. The latter
  are shown to provide the characteristic length scale. The model can
  explain the origin of important features of dunes, such as the
  formation of a slip face, the broken scale invariance, and the
  existence of a minimum dune size. It also predicts the longitudinal
  shape and aspect ratio of dunes and heaps, their migration velocity
  and shape relaxation dynamics. Although the minimal model employs
  non--local expressions for the wind shear stress as well as for the
  sand flux, it is simple enough to serve as a very efficient tool for
  analytical and numerical investigations and to open up the way to
  simulations of large scale desert topographies.
\end{abstract}
 


\maketitle

\section{Introduction}
  
\begin{figure}[t]
  \includegraphics[width=\columnwidth]{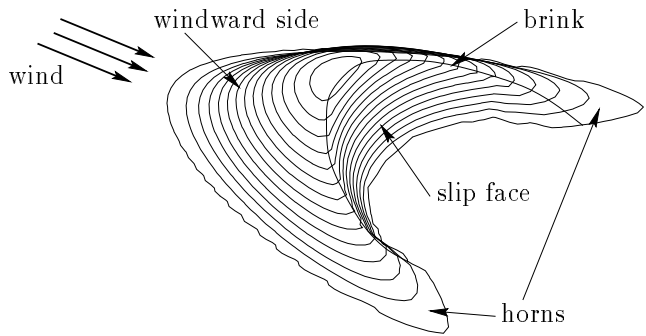} 
\caption{Sketch of a barchan dune. Sand is eroded by the wind on the
upwind or ``stoss'' side and transported to the brink. Strong
deposition occurs due to flow separation behind the brink. On the
downwind or ``lee'' side, sand slides down at the angle of repose
(about $32^{\circ}-35^{\circ}$) over a concave slip face.}
\label{fig:barchan}
\end{figure}

Sand dunes develop wherever sand is exposed to an agitating medium
such as air or water that lifts grains from the ground and entrains
them into a surface flow. The diverse conditions of wind and of sand
supply in different regions on Earth give rise to a large variety of
shapes of aeolian dunes \cite{Bagnold41,Pye90,Lancaster95}. Moreover,
dunes have been found on the sea--bottom and even on Mars
\cite{malin-etal:98,Thomas99}.  Despite the long history of the
subject, the underlying physical mechanisms of dune formation are
still not very well understood.  How are aerodynamics (hydrodynamics)
and the particular properties of granular matter acting together to
create dunes?  How is the shape of a dune maintained when it moves?
Since the macroscopic phenomena of interest are separated by many
orders of magnitude from the grain scale and involve various coupled
nonlinear processes such as turbulent air flow and grain hopping
(``saltation''), one is bound to devise some simplified models in
order to address such questions. We will argue that approximate
numerical models can only be successful if based on a sound
qualitative understanding of the problem. Therefore, our main aim is
to identify the key mechanisms underlying dune formation and migration
and incorporate them into a working minimal model of aeolian sand
dunes, and we will emphasize generic aspects over the more specific
details. For definiteness, the reader may find it helpful to think of
isolated transverse dunes or crescent--shaped barchan dunes as major
applications of the model. A schematic sketch of the height profile of
a barchan is shown in Fig.~\ref{fig:barchan}. The broad phenomenology
of aeolian and submarine land forms provides a large number of
different characteristic structures that can certainly not all be
described by the same simple model developed with the specific
examples of barchan or transverse dunes in mind. However, we expect
that our approach is amenable to future adaptations that make it
applicable to a broader class of sand topographies on the one hand,
and for quantitative investigations of more specific questions on the
other hand. Although the minimal model refers only to rather generic
properties of the wind velocity field and the laws of aeolian sand
transport, it can make interesting predictions about the surface
profile, the development and position of the slip face, dune migration
etc.\ that are insensitive to the simplifying assumptions. The main
features of the model were already briefly presented in a recent
Letter \cite{kroy-sauermann-herrmann:2002}.  The present contribution
gives a more comprehensive discussion of the model and tries to
communicate its precise definition as well as its major predictions to
an interdisciplinary readership. The model, as presented here, is
restricted to a two--dimensional ($2d$) slice of a dune parallel to
the unidirectional wind.  (A generalization to $3d$ problems is in
preparation.) A further restriction is the neglect of ripples and
direct slope effects onto the sand transport outside slip
faces. Although they have successfully been incorporated into
continuum sand transport models
\cite{hoyle-woods:97,hoyle-mehta:99,prigozhin:99} similar to our own
\cite{sauermann-kroy-herrmann:2001}, we chose to disregard them for
the present purpose and leave their integration to future work.

The paper is organized as follows. In the next introductory section we
summarize some background knowledge and basic definitions. We will
also introduce a naive ``zeroth order'' description of the wind shear
stress and the induced aeolian sediment transport. Its instructive
failure to produce dune--like steady--state solutions will be a guide
for identifying two relatively small effects (the upwind shift of the
maximum of the shear stress with respect to the topography and the
saturation transients in the sand flux) as key ingredients of a proper
description of structure formation by aeolian sand transport.  We will
moreover derive the scaling behavior of the migration velocity for
translation invariant heaps and dunes of different size but similar
shape based on very general grounds. Sections~\ref{sec:wind} and
\ref{sec:saltation} are devoted to the definition of the minimal
model, i.e., to the modeling of the air shear stress exerted onto a
heap of sand and the induced sand transport, respectively. The first
step builds on turbulent boundary layer calculations developed in a
series of publications mainly by Hunt and coworkers
\cite{jackson-hunt:75,%
sykes:80,zeman-jensen:88,hunt-leibovich-richards:88,Carruthers90,weng-etal:91},
the second one on a previous contribution
\cite{sauermann-kroy-herrmann:2001} by the present authors. Only the
most pertinent results of these earlier developments will be
summarized here.  In the remainder, we will derive some important
predictions of the model for the central slice of a barchan dune or
transverse dune. In particular, we will demonstrate that there is a
minimal dune size. Although we will thereby gain interesting results,
these are rather meant to be illustrative examples of possible
applications of the model. By no means do we attempt to provide a
complete analysis of its predictions, and it should become obvious
that much more remains still to be done.  Finally we will summarize
our main results and speculate about probable consequences of the
present $2d$ theory for $3d$ topographies.

\section{General}\label{sec:general}
\subsection{Aeolian sand transport}\label{sec:aeolian}
Before going into the description of the model, we want to recall some
general background and to introduce some quantities of major interest.
First of all, for convenience, we will usually refer to dunes without
slip face as heaps. Further, we will sometimes find it helpful to
focus on isolated heaps or dunes on bedrock, although most of our
discussion is not restricted to this situation.  

The key quantity for the description of the formation and migration of
sand dunes and heaps is the local horizontal surface velocity $v(x,t)$
of a sand height profile $h(x,t)$ at all positions $x$ and times
$t$. Via mass conservation it can be related to the erosion rate
$\nabla q(x,t)$ (negative erosion is deposition), where the sand flux
$q(x,t)$ is defined as the mass of sand transported per unit of time
across a hyper--plane transverse to the wind direction. More
precisely, since we want to specialize our discussion to a $2d$ slice
parallel to the unidirectional wind velocity, the hyperplane is a
vertical line and $q$ is a flux per unit width. Mass conservation then
takes the form of a continuity equation for the height profile
\begin{equation}\label{eq:mass}
\varrho_{s}\frac{d h(x,t)}{dt} = -\frac{\partial q(x,t)}{\partial x} 
\end{equation}
with $\varrho_{s}$ the density of the sand bed. 

With Eq.(\ref{eq:mass}) one can write the position dependent migration
velocity at a given time $t$ as
\begin{equation}\label{eq:v_def}
v(x) = \varrho_{s}^{-1} \; \frac{q'}{h'}
\;,
\end{equation}
where we have introduced the shorthand notation $f'(x)\equiv
df(x)/dx$.  At this stage we can already get some physical insight by
observing that this equation needs special attention at the top of a
heap or dune, where we expect the denominator to vanish. For $v$ to
remain finite at the crest as required in the steady state, there are
in general only two possibilities. Either the sand flux $q$ is fine
tuned so that the erosion $q'$ vanishes in exactly the same way as the
slope $h'$, or the profile $h(x)$ is not differentiable at the
crest. As the reader may already anticipate and will be verified
below, both cases have their physical realizations, the former
in heaps or small dunes with smooth crests and the latter in large
dunes with a slip face that terminates in a sharp brink.

The problem we face, if we want to calculate the dynamic evolution of
desert topographies, is the closure of Eq.(\ref{eq:mass}) or
(\ref{eq:v_def}) by expressing the flux $q(x,t)$ in terms of the
height profile $h(x,t)$ and the external wind and boundary
conditions. Since for the applications we have in mind, the migration
velocity is very small compared to the speed of elementary sand
transport processes (grain hopping etc.) and the wind speed, the
topography can be assumed to be stationary for considerations
concerning the wind and sand transport dynamics. This allows one to
subdivide the problem of calculating $q(x)$ into two independent
steps. First, one needs to know the stationary wind velocity above a
given topography. More precisely, what is required is the shear stress
$\tau$ exerted by the wind onto the ground. And secondly, one needs a
model that predicts the stationary sand flux $q(x)$ for a given
stationary $\tau(x)$, schematically
\begin{align}
h(x) &\to \tau(x) \label{eq:task1} \\
\tau(x) &\to q(x)\;. 
\label{eq:task2}
\end{align} 
Computing the derivative $q'$ and integrating the mass conservation
Eq.(\ref{eq:mass}) then closes the model and allows one to predict the
development of the surface profile in time.  Since aeolian dunes
typically have relatively gentle slopes outside their slip face, we
will at the present stage restrict the scope of the minimal model to
this case and disregard in Eq.(\ref{eq:task2}) the direct slope
effects $h'(x) \to q(x)$ onto the flux outside the slip face.

In special cases, the relations (\ref{eq:task1}), (\ref{eq:task2}) are
phenomenologically and theoretically well established.  For a flat
surface, $h(x)\equiv$ const., it is well known
\cite{landau-lifshitz:fm} that the mean turbulent wind velocity
increases logarithmically with height above the surface. It can be
characterized by a single characteristic velocity, the ``shear
velocity'' $u_*$ defined by $u_*^2\equiv\tau_0/\varrho_{a}$ with
$\tau_0$ the (suitably time averaged) shear stress and $\varrho_{a}$
the density of air. Since the shear stress of the air is transmitted
to the surface, the latter can mobilize grains on a surface covered
with sand, if it exceeds a threshold value $\tau_t$. As a result, the
wind entrains some grains into a surface layer flow.  The grains
advance mainly by an irregular hopping process (``saltation''),
thereby reducing the wind velocity in the surface layer. Via this
feed--back mechanism a unique relation between the shear stress $\tau$
and the sand flux $q$ is established in the equilibrium state.  If
$\tau$ is not too close to the threshold, this relation can
approximately be represented as \cite{Bagnold41}
\begin{equation}\label{eq:bagnold} 
q_s \propto \tau^{3/2} \;.     
\end{equation} 
Although a host of more accurate descriptions have been discussed in
the literature
\cite{Lettau78,Sorensen91,Pye90,ActaMechanica91,sauermann-kroy-herrmann:2001}
and one of them will be part of our definition of the minimal model
below, the simpler Eq.(\ref{eq:bagnold}) will be sufficient for our
qualitative discussion in the first part of the paper. The index $s$
in Eq.(\ref{eq:bagnold}) emphasizes that such local relations are
restricted to situations where the flux is saturated, that is, equal
to its equilibrium transport capacity. This is certainly not the case
near a boundary between uncovered and covered ground or on sloped
beds. Neglecting this restriction for the moment,
Eq.(\ref{eq:bagnold}) predicts that the shear stress perturbation
\begin{equation}\label{eq:tauhat} 
  \hat\tau(x) \equiv \tau(x)/\tau_0-1
\end{equation} 
above a modulated topography $h(x)$ is responsible for flux gradients
$dq_s/dx$ that cause erosion and deposition and thus --- according to
Eqs.(\ref{eq:mass}), (\ref{eq:v_def}) --- migration of the sand surface.
Explicitly closing the model by assuming that the shear stress is an
affine function of the modulation of the topography ($\hat \tau
\propto h$) leads to what we call the ``zeroth order'' model, which
will briefly be analyzed in the next paragraph.

\subsection{The ``zeroth order'' model}\label{sec:zero}
The zeroth order model is given by
\begin{align}
\hat\tau \{h(x)\} & \to \hat\tau(h)\propto h(x)/L \label{eq:tau0} \\ 
q\{\tau(x)\}  & \to q(\tau)= q_s(\tau) \label{eq:q0}
\end{align}
where we have used the curly brackets to indicate a general functional
dependence and introduced a characteristic length scale $L$ of the
topography to normalize the height profile. (The motivation for the
latter step will become clear in the next section.) The zeroth order
model assumes \emph{local} relations in
Eqs.(\ref{eq:task1}),(\ref{eq:task2}). It approximates the wind shear
stress perturbation by its ``affine'' contribution (proportional to
the profile $h$ that causes the perturbation) and replaces the true
sand flux $q$ by its saturated value $q_s$, thereby neglecting
saturation transients.  This model is so simple that its qualitative
predictions for an arbitrary smooth heap of sand can easily be
anticipated without doing any actual calculations. 

Combining Eq.(\ref{eq:v_def}) with
Eqs.(\ref{eq:bagnold})--(\ref{eq:q0}) one obtains a surface velocity
that increases with height ($dv/dh \geq 0$) due to the nonlinearity
of Eq.(\ref{eq:bagnold}). This implies that the upwind (or ``stoss'')
slope tends to decrease and the downwind (or ``lee'') slope tends to
increase. Since $dq/dx \propto dh/dx$ by the chain rule, there is no
erosion or deposition at the top of a smooth heap, which therefore
keeps its initial height. Obviously, integrating forward in time will
eventually increase the lee slope up to the angle of repose, where
surface avalanches have to be introduced and a slip face of constant
slope develops.  If the latter reaches the crest the above argument
for the persistence of the height can no longer be applied, because
the slope at the crest is then ill defined. Since there is so far
nothing to stop a further decrease of the windward slope, the model
dune will then start to decrease in height and finally flatten
out. The steady--state solution is a flat surface.

The simple argument shows that the zeroth order model --- although it
gives some clue as to the origin of the slip face --- is insufficient
for a proper qualitative understanding of dunes. However, some
important lessons can be learned from it that will be helpful in our
further investigation of the problem. First, even with a very
simplistic model any reasonably heap--like initial condition will
quickly develop into a dune--like shape with a slip face. Secondly,
although the latter may seem to converge to a steady--state solution
for intermediate times, it finally turns out to be unstable and
flattens out. The discussion of the migration velocity in
Section~\ref{sec:aeolian} suggests that small deviations from
Eqs.(\ref{eq:tau0}) and (\ref{eq:q0}) at the brink can make an
important difference. Obviously some caution is needed in judging the
success of numerical models of dune formation. Unless stability has
explicitly been demonstrated, they may be suspected to fail in a
similar way as the zeroth order model when integrated over
sufficiently long times (which has actually not been checked for some
models that can be found in the literature) or to be sensitive to
numerical errors at the brink. Detailed numerical modeling should
therefore be preceded by a sound qualitative understanding of the
mechanisms underlying dune formation.  We will argue in
Sections~\ref{sec:wind}, \ref{sec:saltation} that to this end a subtle
balance between two small deviations from
Eqs.(\ref{eq:tau0}),(\ref{eq:q0}) and especially non--local
contributions in Eq.~(\ref{eq:task2}) have to be taken into account.

\subsection{Migration velocity}\label{sec:v}
Before entering a detailed discussion of the minimal model, it is
worth pausing for some general thoughts as to what can be said about
the shear stress and the speed--up of the wind above an obstacle,
without actually doing the (somewhat involved) calculation. 

\begin{figure}[t]
\psfrag{H}{$H$}
\psfrag{L}{$L$}
\psfrag{e}{$\varepsilon$}
\psfrag{u}{$u$}
\psfrag{ee}{$u+\varepsilon u $}
\begin{center}
\includegraphics[width=\columnwidth]{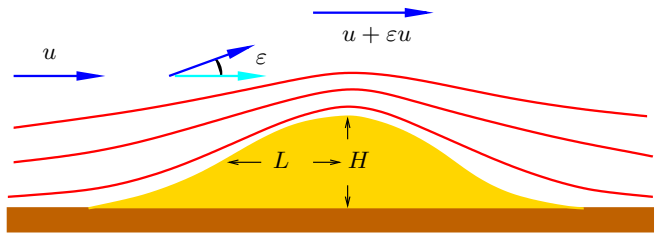}
\end{center}
\caption{Schematic sketch of the deflection of the wind velocity $u$
above a flat heap of aspect ratio $\varepsilon \equiv H\!/\!
L\ll1$. The characteristic length scale $L$ is in this context
conventionally often identified with  the half length at half
height of the heap. The vertical deflection causes a speed--up above
the top of the heap. This is accompanied by a pressure perturbation
that is negative above the top of the heap and positive at its tails.
Due to turbulence, the flow pattern is asymmetric even above a
symmetric heap.}
\label{fig:scaling}
\end{figure}

A basic property of strongly developed turbulence is its dilation
invariance or scale free structure. Whereas general Navier--Stokes
flow is invariant under a scale transformation that keeps the Reynolds
number constant, strongly turbulent flow (for ``infinite'' Reynolds
number) allows for infinitely many such similarity
transformations. Landau and Lifshitz \cite{landau-lifshitz:fm} took
advantage of this fact for deriving the logarithmic velocity profile
mentioned above by an elegant scaling argument. The logarithmic
velocity profile suggests that the speed--up of the wind and therefore
also the shear stress perturbation above a heap of given shape should
itself be logarithmically dependent on its size. But how do they
depend on the shape of the obstacle?  Since the flow itself does not
provide any characteristic length scale, the dimensionless quantity
$\hat \tau$ defined in Eq.(\ref{eq:tauhat}) can only depend on a
dimensionless characterization of the profile $h(x)$. In other words,
to lowest order in the perturbation, it must be a linear functional of
the derivative $h'$ and can be written as
\begin{equation}\label{eq:scaling}
\hat \tau(\xi) = 
	\varepsilon \,{\cal T}\left\{f'(\xi)\right\}\;, \qquad
	\varepsilon \equiv H/\! L \;,
\end{equation}
with a dimensionless profile function
\begin{equation}\label{eq:profile}
f(\xi)\equiv h(x)/H \qquad \xi \equiv x/\! L \;. 
\end{equation}
and a scale--free (and necessarily non--local) linear functional
${\cal T}$.  This reasoning can be repeated for the dimensionless
velocity and pressure perturbations.  Intuitively, the scaling $\hat
\tau \propto \varepsilon$ for a flat smooth obstacle ($\varepsilon \ll
1$) can be understood from Fig.~\ref{fig:scaling}. When the air flows
over the obstacle, the velocity close to the obstacle is deflected by
an angle $\varepsilon$ whereas it remains constant far above the
obstacle. For incompressible flow, continuity translates this into a
speed--up of order $\varepsilon$ and (via Bernoulli's law) into a
corresponding pressure drop near the top of the heap. This in turn
causes a shear stress perturbation $\hat \tau$ of the same order.

These general considerations already allow us to predict the scaling
of the migration velocity $v$ with dune size if we assume that dunes
of different size have roughly similar shapes $f(\xi)$ and aspect ratios
$\varepsilon$, which is indeed suggested by the scale invariance of
the turbulent wind field and by observations.  Inserting
Eq.(\ref{eq:scaling}) into Eq.(\ref{eq:v_def}), and again
approximating Eq.(\ref{eq:task2}) by a local sand transport law $q
\equiv q(\tau)$, we find
\begin{equation}\label{eq:v}
 v  \frac{dq}{d\hat\tau}
	\frac{d\hat\tau/dx}{\varrho_{s}\varepsilon f'}
	=\frac{dq}{d\hat\tau}\frac{{\cal
	T'}\{f'\}}{\varrho_{s} L f'} \propto \frac1L \;.
\end{equation}
The final proportionality strictly holds only if the steady--state
shape $f(\xi)$ and aspect ratio $\varepsilon$ are scale
invariant. However, it can be expected to be robust and rather
insensitive against violations of exact scale invariance.  First, the
normalized steady--state shapes $f(\xi)$ are strongly constrained by
the requirement that they render $v(x)\equiv v$, independent of $x$,
along the heap. Therefore, they should to a first approximation be
independent of size, which is indeed borne out by the minimal model
(Fig.~\ref{fig:shapes}) and empirical observations
\cite{sauermann-etal:2000}.  Moreover, the dependence $v\{ f'(\xi) \}$
is rather indirect and can therefore be expected to be weak. Secondly,
Eq.(\ref{eq:bagnold}) suggests that for gently sloped obstacles
($\varepsilon \ll 1$) the dependence of $dq/d\tau$ on the aspect ratio
$\varepsilon$ also is not very pronounced. And finally, --- due to the
above mentioned scale invariance of turbulence --- a scale invariant
aspect ratio can reasonably be expected for large dunes. In fact, we
will show below that the minimal model predicts that the aspect ratio
of small heaps is not constant but rather decreases proportional to
their height. But this also implies that the latter becomes too small
to have a very significant effect on the above argument.  Note,
however, that only for strictly scale invariant dunes, Eq.(\ref{eq:v})
becomes identical to the often quoted observation that dunes migrate
with a speed inversely proportional to their height
\cite{speed}. Since the deviations of large dunes from scale
invariance are not very pronounced, the difference between these
predictions is not very strong except for small dunes and heaps.
Presently available field data are maybe not accurate enough to
clearly distinguish between $v\propto 1/L$ and $v\propto 1/H$, though
some data support $v\propto1/L$, most notably the comprehensive study
of barchan dunes in southern Peru by Finkel \cite{Finkel59}. As we
will show below, our numerical results for the minimal model clearly
favor $v\propto1/L$.

We also note that together with Eq.(\ref{eq:bagnold}), Eq.(\ref{eq:v})
moreover predicts that the migration velocity grows non--linearly with
(as the third power of) the wind velocity. A more accurate relation
can be obtained from the minimal model as described below, but the
qualitative conclusion is the same. Dunes can migrate farther in a
short period of exceptionally strong wind than during much longer
periods of gentle winds. Finally, we should mention that some caution
is needed when identifying the characteristic length scale $L$ in
Eq.(\ref{eq:v}). In our discussion, we have so far assumed that
$f(\xi)$ is a smooth function, which is not the case for dunes with
slip face. Below we will argue that in this case $f$ should be
identified with the envelope of the dune and its separation bubble and
$L$ with the characteristic length scale of this envelope. For a
barchan dune the latter practically coincides with the total length of
the dune from its windward end to the tips of its horns (cf.\
Fig.~\ref{fig:barchan}).

In contrast to the overall migration velocity of a translation
invariant dune, the position dependent migration velocity $v(x)$ that
determines that shape is much harder to obtain since it requires a
precise knowledge of the non--local functional ${\cal T}$ in
Eq.(\ref{eq:scaling}). This will be provided in the following section.

\section{Wind shear stress}\label{sec:wind}
\subsection{Surface shear stress on a smooth heap}\label{sec:theory}
\begin{figure}[tb]
   \psfrag{L}{$10^{-5} L/z_0$}
   \psfrag{A}{$A$, $B$} 
\begin{center}  
   \includegraphics[width=\columnwidth]{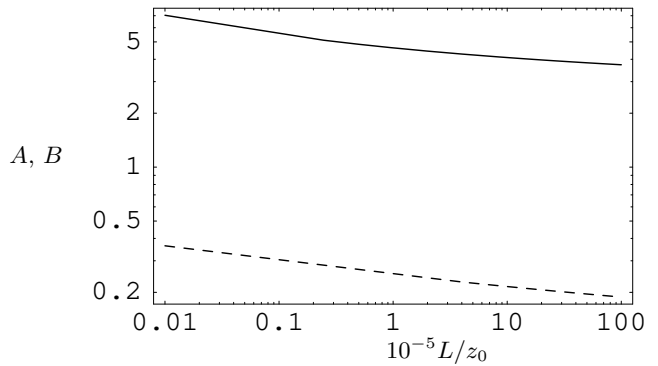} 
\caption{The
   theoretical prediction for the dependence of the parameters $A$
   (solid line) and $B$ (dashed line) of Eq.(\ref{eq:a_b}) on the
   ratio of the dune size $L$ to the roughness length $z_0$.}  
\label{fig:ab}
\end{center}
\end{figure}

The discussion in the preceding paragraph showed that --- in contrast
to the assumption in Eq.(\ref{eq:tau0}) --- the dependence of the
shear stress on the height profile is non--local. Although it will
turn out that this shortcoming of Eq.(\ref{eq:tau0}) is not
responsible for the failure of the zeroth order model, it should by
now have become apparent that further progress can hardly be achieved
without a rather detailed understanding of the turbulent wind field
above heaps and dunes. For dunes with a slip face that typically has a
slope of about $32^{\circ}-35^{\circ}$ and terminates in a sharp
brink, the situation is similar to the textbook example of a backward
facing step, which has the reputation of a test case for numerical
turbulent models. Even if a commercial turbulent solver is used, the
accurate calculation of the shear stress e.g.\ on a barchan dune is a
non--trivial task and quite demanding in computer time and memory, and
the most interesting long--time dynamics of dunes is therefore
difficult to access. For this reason, we want to focus on flat smooth
heaps, first. In this case, one can apply an analytical perturbation
theory for turbulent boundary layer flow over smooth hills that has
been developed over the last decades \cite{jackson-hunt:75,sykes:80,%
zeman-jensen:88,hunt-leibovich-richards:88,Carruthers90,weng-etal:91}.
Though the calculation is essentially a formalization of the intuitive
description accompanying Fig.~\ref{fig:scaling}, it requires a highly
non--trivial boundary layer construction that we will not recapitulate
here. The interested reader is referred to the original literature. We
merely quote the final result for the $x-$component (along the main
wind direction) $\hat \tau_x$ of the surface shear stress perturbation
above a profile $h(x,y)$
\cite{hunt-leibovich-richards:88,weng-etal:91},
\begin{equation}
\label{eq:hunt} 
{\cal F}_{xy}\{\hat\tau_x\} = \frac{A k_x( k_x +
i B |k_x|)}{(k_x^2 + k_y^2)^{1/2}} \, {\cal F}_{xy} \{ h(x,y) \}
\;.  
\end{equation} 
We have abbreviated the Fourier transformation from the space
variables $x$, $y$ to the respective wave numbers $k_x$, $k_y$ by
${\cal F}_{xy}$. For simplicity the logarithmic $k-$dependence of the
parameters $A$ and $B$ was neglected.  The latter are then given
by
\begin{equation}\label{eq:a_b}
\begin{split}
 A &= 
\frac{\ln \left(\Phi^2/\ln \Phi\right)^2}{2(\ln \phi)^3}
 [1+\ln \phi+2\ln(\pi/2)+4\gamma] \\
 B &= \pi/[1+\ln\phi+2\ln(\pi/2)+4\gamma_E] \\
  \phi & \equiv 2\kappa^2\Phi/\ln\phi 
\end{split} 
\end{equation}
and depend logarithmically on the ratio $\Phi\equiv L/z_0$, where $L$
is the characteristic length of $h(x,0)$ (for this purpose
conventionally often identified with half the length at half height or
about one fourth of the characteristic wavelength) and $z_0$ is a
measure of the surface roughness (typically an effective length
somewhat below the linear dimension of the latter). We also have
introduced the von K\'arm\'an constant $\kappa\approx 0.4$ and Euler's
constant $\gamma_E\approx 0.577$.  A practical approximation for
$\phi$ is obtained by iterating (twice) the implicit equation for
$\phi$ and closing it by dropping the remaining $\ln\phi$.  The
dependence of $A$ and $B$ on $L$ is depicted in
Fig.~\ref{fig:ab}. Obviously, as long as $L/z_0$ does not change by
orders of magnitude (e.g.\ due to vegetation), this extremely weak
scale dependence is negligible for our purposes, and $A$ and $B$ can
be regarded as constant theoretical or phenomenological parameters.
For definiteness we often work with the values $A=4$ and $B=0.25$
(approximately obtained for $L/z_0 = 10^5 \dots 10^6$). Although these
values may differ somewhat depending on the particular application in
mind, or on the presence or absence of ripples, and may
phenomenologically be somewhat different from the theoretical
prediction, this does not affect our general conclusions.

For the following discussion we want to specialize Eq.(\ref{eq:hunt})
to the central slice of a transverse or barchan dune along the wind
direction. To this end we evaluate Eq.(\ref{eq:hunt}) for the central
slice $h(x)$ of a heap that has a Gaussian shape with standard
deviation $\sigma$ in the transverse direction parameterized by $y$,
\begin{equation}\label{eq:gauss}
h(x,y)=h(x)e^{-y^2/2\sigma^2} \;.
\end{equation}
This approximation is technically useful, and although it may seem
relatively crude for a particular real dune, it typically does not
introduce any noteworthy derogation of the results compared to a more
accurate description. The Fourier coefficients of the shear stress
$\hat\tau(x)\equiv \hat\tau_x(x,0)$ on the central slice along the
wind direction become
\begin{multline}
{\cal F}\{\hat\tau(x)\}= \frac{\sigma A}{\sqrt{2\pi}} k(k +
iB |k|) \times
\\
 e^{-\frac14k^2\sigma^2}
K_0\!\left(\frac{k^2\sigma^2}4\right){\cal F}\{h(x)\}
\end{multline}
Here, $K_0$ denotes a modified Bessel function and the Fourier
transforms are one--dimensional, so that we can drop the redundant
$x-$subscripts.  

\begin{figure}[t]
\psfrag{x}{$x/L$}
\psfrag{De}{$\Delta_1$} 
\psfrag{Ds}{$\Delta_2$} 
\begin{center}
 
\includegraphics[width=\columnwidth]{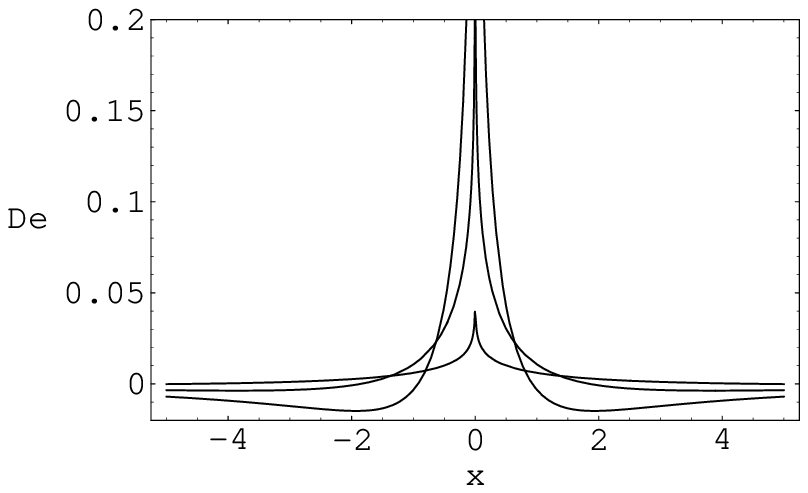} 
\includegraphics[width=\columnwidth]{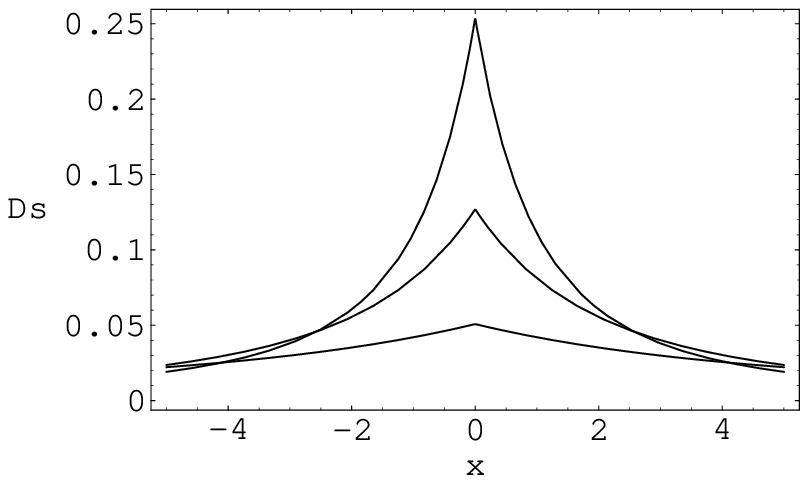}
\caption{The peaked functions $\Delta_1$ and $\Delta_2$ of
Eq.(\ref{eq:deltas}) for $\sigma=1$, 2, 5. The area under the peaks
remains constant (0 and $1/2\pi$), while the peak heights decrease
proportional to $\sigma^{-2}$ and $\sigma^{-1}$, respectively.}
\label{fig:corrections} 
\end{center} 
\end{figure}

For transverse dunes ($\sigma/L\to \infty$), we obtain the two
equivalent expressions
\begin{subequations}\label{eq:wind}
\begin{align}
{\cal F}\{\hat\tau^\infty (x)\} &= A(|k|+iB\, k){\cal F}\{h(x)\} \;. 
\label{eq:fourier_wind} \\
 \hat\tau^\infty(x) &= A \left[ h'(x)\otimes (\pi x)^{-1} + B h'(x)\right]
 \;. \label{eq:real_wind}
\end{align}
\end{subequations}
For the real space version we have abbreviated a convolution integral
according to
\begin{equation}
\label{eq:convolution}
f\otimes g \equiv \int_{-\infty}^{\infty}\!\!\!\!  d\xi \; f(\xi)\,g(x-\xi) \;.
\end{equation} 
Evaluation for arbitrary $\sigma$ gives two correction terms
\begin{equation}
\label{eq:correction}
\hat\tau^\sigma=\hat\tau^\infty - A (h\otimes \Delta_1 + B
h'\otimes \Delta_2) \;,
\end{equation}
with 
\begin{align}
\label{eq:deltas}
\Delta_1 &=
\frac{U\left(\frac32,1,\frac{x^2}{2\sigma^2}\right)}{4\sqrt \pi
\sigma^2}+\frac{\frac{\sqrt{\pi}}2
U\left(\frac12,0,\frac{x^2}{2\sigma^2}\right)-1}{\pi x^2} \\ 
\Delta_2 &=\frac1{\sigma \pi}\int_{0}^{\infty}
\!\!\!\!d\xi\,\cos\left(\frac{\xi x}\sigma\right) \left[1-\frac{\xi
e^{\xi^2/4}}{\sqrt{2\pi}} K_0\left(\frac{\xi^2}4\right)\right]
\end{align}
two even functions depicted in Fig.~\ref{fig:corrections} that are
flat for $\sigma/L\to\infty$ and become peaked for $\sigma\simeq
L$. (The confluent hypergeometric $U$ functions
\cite{abramowitz-stegun} have been introduced to rephrase the
sine--part of the Fourier integrals.)

\begin{figure}[t]
\psfrag{x}{$x/L$}
\psfrag{ta}{$\hat \tau(x)$} 
\begin{center} 
\includegraphics[width=\columnwidth]{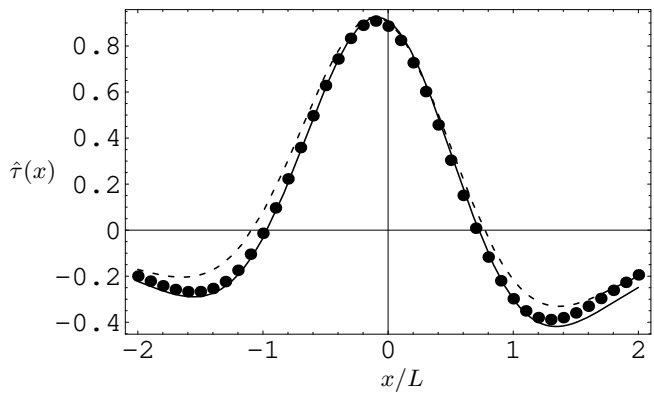} 
\caption{Shear stress perturbation above the central slice of a $3d$
symmetric ($\sigma=L/\sqrt 2$) Gaussian heap. The plot compares two
approximations to Eqs.(\ref{eq:hunt}), (\ref{eq:correction}) (points):
(i) Eq.(\ref{eq:approxcorrect}) with $Lc_1\approx c_2\approx0.2$
(solid line), and (ii) Eq.(\ref{eq:wind}) with $A$ renormalized by a
factor 0.8 (dashed line). While (i) is practically indistinguishable
of Eq.(\ref{eq:hunt}) on the present level of accuracy, the simpler
approximation (ii) already captures most of the $3d$ effects.}
\label{fig:hunt_jackson} 
\end{center} 
\end{figure}

Since the correction terms in Eq.(\ref{eq:correction}) are numerically
small, we may --- given a reasonable localized heap shape in the wind
direction --- approximately replace both functions $\Delta_1$ and
$\Delta_2$ by delta functions, thus arriving at
\begin{equation}\label{eq:approxcorrect}
\hat\tau_\sigma\approx \hat\tau_\infty - A [c_1(\sigma) h + B
c_2(\sigma) h'] \;.
\end{equation}
In this approximation they are seen to give merely a
$\sigma-$dependent renormalization of the asymmetry parameter $B\to
B(\sigma) \lesssim B(\infty)= B$ and to add a (trivial) term
$c_1(\sigma) h(x)$ within the brackets of
Eq.(\ref{eq:real_wind}). Numerically, one can estimate $L\,c_1(
L/\sqrt2)$ and $c_2( L/\sqrt2)$ to be about 0.2 (cf.\
Fig.~\ref{fig:hunt_jackson}). The exact $\sigma-$dependence of the
coefficients is determined by the shape and extension of the heap in
the $x-$direction, because the area under the peaks $\Delta_1$ and
$\Delta_2$ is constant and independent of $\sigma$, while the peak
height decreases proportional to $\sigma^{-2}$ and $\sigma^{-1}$,
respectively. Since both corrections vanish for $\sigma/L\to\infty$
and do not contribute any substantial new effects to
Eq.(\ref{eq:wind}), they may for simplicity be omitted altogether in
the following discussion that mainly aims at a qualitative
understanding.  Fig.~\ref{fig:hunt_jackson} moreover shows that they
can approximately be mimicked by a renormalized parameter $A$ in
Eq.(\ref{eq:wind}) for the central slice of a symmetric heap. This
leads to the important conclusion that the wind shear stress on the
central slice of a $3d$ symmetric heap and on a heap with a profile
that is constant in the transverse direction, are qualitatively the
same and quantitatively similar, which was not \emph{a priori}
obvious.  Together with the fact that on a gently sloped obstacle the
transverse components of the shear stress are small compared to its
longitudinal components, this suggests that the predictions of
Eq.(\ref{eq:wind}) apply in a first approximation to any slice of a
dune parallel to the wind direction. In this sense, the study of
Eq.(\ref{eq:wind}) is representative.  Summarizing the foregoing
discussion we can say that to gain a qualitative understanding of dune
formation by aeolian sand transport one may focus on
Eq.(\ref{eq:wind}) as a model for the wind shear stress. We therefore
analyze this equation in some detail in the next paragraph.

\subsection{Properties and consequences of Eq.(\ref{eq:wind})}\label{sec:properties}

A scaling analysis of Eq.(\ref{eq:wind}) immediately reveals
that $\hat \tau$ is indeed of the general form anticipated on general
grounds in Eq.(\ref{eq:scaling}). The amplification of the shear
stress at the top of a smooth profile is thus determined by its aspect
ratio $\varepsilon=H\!/\!L$ and is essentially independent of the
absolute height $H$. It only has a very weak logarithmic dependence on
the absolute size of the dune through the prefactors $A$ and $B$ given
in Eq.(\ref{eq:a_b}). Moreover, for a symmetric profile
$f(-\xi)=f(\xi)$, $\hat \tau$ is the sum of a symmetric part and an
anti--symmetric part, i.e., the flow over the heap has a symmetry
breaking component that is a consequence of turbulence.  The origin of
the symmetric and antisymmetric parts of $\hat \tau$ can intuitively
be understood as follows. As we have pointed out in
Section~\ref{sec:v} (see Fig.~\ref{fig:scaling}), the streamlines have
to be compressed above the heap if the perturbation is not to be
transmitted to infinite height, and as a consequence, there is a
corresponding increase in the shear stress. For the laminar average
flow, this speed--up and the associated decrease in atmospheric
pressure above the heap are symmetric for a symmetric heap as is the
corresponding shear stress perturbation, which accounts for the
dominant symmetric part of $\hat \tau$. On the other hand, the inertia
of the turbulent velocity fluctuations around this laminar main flow
contributes an asymmetric resistance to deflections of the flow. It
counteracts the upturn of the streamlines on the windward side and the
downturn on the lee side. Formally, this effect enters the
perturbative calculation of $\hat \tau$ through the Reynolds stress.

\begin{figure}[t]
\psfrag{x}{$x$}
\psfrag{h}{\mbox{}\;\; $f$} 
\psfrag{t}{$\!\!\!\!\tau/\tau_0$} 
\begin{center}
 
\includegraphics[width=\columnwidth]{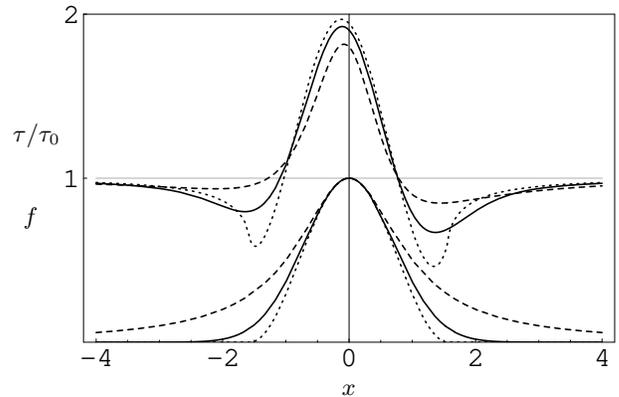} 
\caption{\emph{Lower curves:} The normalized profiles of
Eq.(\ref{eq:hills}), the Lorentzian $f_L$ (dashed), Gaussian $f_G$
(solid), and cosine $f^2_C$ (dotted). \emph{Upper curves:} The
corresponding surface shear stresses $\tau(x)/\tau_0$ from
Eq.(\ref{eq:solutions}) with $A \varepsilon = 0.8$, $B=0.25$.}
\label{fig:hills} 
\end{center} 
\end{figure}

Further insight can be gained from special analytical
solutions to Eq.(\ref{eq:wind}).  For the normalized heap
profiles
\begin{equation}
   \label{eq:hills}
\begin{split}
   f_L(\xi) & = \frac1{1+\xi^2}\\
   f_G(\xi) & = \exp\left(-\xi^2\right)\\ 
   f^n_C(\xi) & = S(\xi) \cos^n \xi  
\end{split}
\end{equation}
with
\begin{equation}
    S(\xi)\equiv  
   \begin{cases}
     1 & |\xi| \leq \pi/2 \\
     0 & |\xi| \geq \pi/2   \;, 
   \end{cases} 
\end{equation}
 we obtain
\begin{equation}
\label{eq:solutions}
\begin{split}
{\hat\tau_L}  
&= A (1 - 2B \xi -\xi^2)f_L^2(\xi) \\ 
{\hat\tau_G}
&=2A \left[\pi^{-1/2} -  \xi(B + \text{erfi}\, \xi) 
f_G(\xi)\right] \\
{\hat\tau_C}
 &=\frac{A}{\pi} \cos(2\xi) \left[\,\text{Si}(\pi+2\xi) +
\text{Si}(\pi-2\xi)\right] \\ & \quad
-\frac{A}{2\pi}
\sin(2\xi)\left[\,\text{Ci}(\pi+2\xi)+\text{Ci}(-\pi-2\xi) \right.  \\ 
& \quad \phantom{-A\sin(2x)[}\left.
-\text{Ci}(\pi-2x) - \text{Ci}(-\pi+2x) \right] \\
& \quad  -2B \, S(\xi)\,\cos \xi\,\sin \xi \qquad (n=2)
\end{split}
\end{equation}
with erfi the imaginary error function and Si and Ci the sine and
cosine integral functions, respectively \cite{abramowitz-stegun}. The
result given for $\hat \tau_C$ is for the special case $n=2$.  Both
the profiles of Eq.(\ref{eq:hills}) and the corresponding solutions of
Eq.(\ref{eq:wind}) given in Eq.(\ref{eq:solutions}) are shown in
Fig.~\ref{fig:hills}.

\begin{figure}[tb]
    \psfrag{X}{$\xi$}
    \psfrag{f}{$f_G$} 
    \psfrag{t1}{$\hat \tau_{\rm sym}$}
    \psfrag{t2}{$\hat \tau_{\rm asy}$} 
    \psfrag{t3}{$\hat \tau_G$}
\begin{center}  
    \includegraphics[width=\columnwidth]{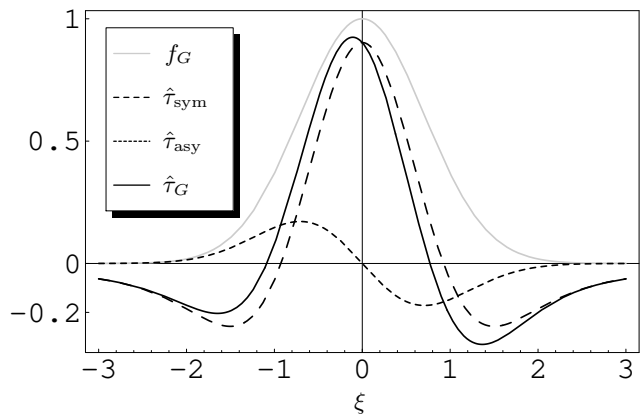}
    \caption{The symmetric and asymmetric parts $\hat \tau_{\rm sym}$
    and $\hat \tau_{\rm asy}$ of the shear stress perturbation $\hat
    \tau_G$ of Eq.(\ref{eq:solutions}) for the Gaussian profile $f_G$
    of Eq.(\ref{eq:hills}) with $A\varepsilon=0.8$ and $B=0.25$. Note
    the small windward shift of the maximum of the shear stress with
    respect to the crest of the heap caused by the asymmetric
    contribution proportional to $B$.}
\label{fig:sym_asym}
\end{center}
\end{figure}

The plots of $\tau$ show that as a rule of thumb one can estimate the
relative magnitude of the shear stress perturbation at the top of the
heap by $A\varepsilon$. The plots share several crucial properties not
present in the affine approximation $\hat\tau\propto h$ of the zeroth
order model. At the tails of the profiles in Fig.~\ref{fig:hills}, the
shear stress decreases below its asymptotic value $\tau_0$ on the
plane.  This effect is particularly pronounced for the profile $f^n_C$
that has a discontinuity in its second derivative.  Further, as a
consequence of the second term in Eq.(\ref{eq:wind}), the surface
shear stress is not symmetric even for symmetric profiles like those
in Eq.(\ref{eq:hills}). Fig.~\ref{fig:sym_asym} displays the symmetric
and asymmetric parts $\hat \tau_{\rm sym}$ and $\hat \tau_{\rm asy}$
of the shear stress perturbation $\hat \tau_G$ for the Gaussian
profile $f_G$, separately. The asymmetric contribution to $\hat \tau$
is small compared to the symmetric one. For the profile $f_L$ the
corresponding shift $\delta x_\tau$ of the location of the maximum of
the shear stress with respect to that of the maximum of $f_L(x)$ can
be calculated analytically,
\begin{equation}\label{eq:shift}
\delta x_\tau/\! L=2\,(1 + B^2)^{1/2}\,\sin [\arctan (B)/3] -B
\;.
\end{equation}
It is indeed found to be very small, because $B$ is small and thus
$\delta x_\tau/\!  L\approx -B/3$ typically amounts to a length of
about a few percent of the total heap length. Nevertheless, it is a
crucial element in the modeling of aeolian sand transport, as will now
be demonstrated.

\begin{figure}[t]
  \psfrag{v}[lr]{$v(x)$} \psfrag{x}{$x/L$}
\begin{center}   
  \includegraphics[width=\columnwidth]{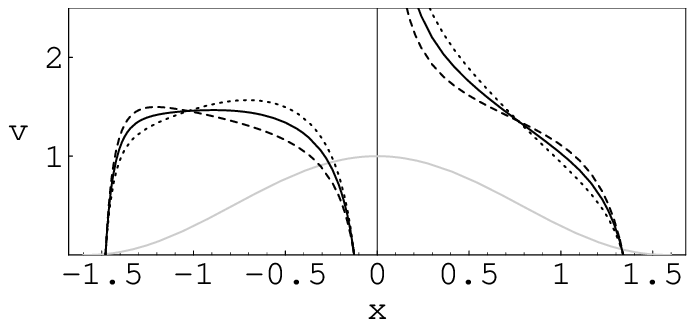}
  \includegraphics[width=\columnwidth]{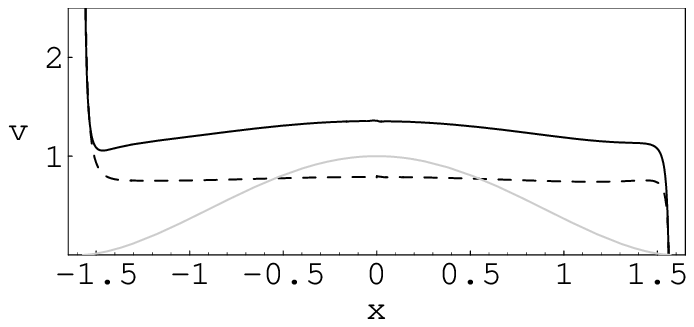} 
\caption{The position
  dependent surface migration velocity $v(x)$ in arbitrary units
  according to Eqs.(\ref{eq:v_def}), (\ref{eq:wind}) with $A=4$,
  $B=0.25$. \emph{Upper part:} For the profile $f^2_C$ (gray) and
  varying aspect ratios $\varepsilon = 0.01$ (dashed), $0.1$ (solid),
  $0.19$ (dotted) and the local flux relation Eq.(\ref{eq:q0}).
  \emph{Lower part:} For the profile $f^{1.65}_C$ (gray) with
  $\varepsilon=0.1$ (solid), $0.05$ (dashed) and the non--local flux
  relation Eq.(\ref{eq:flux}). For simplicity, $q_s$ was represented
  by Eq.(\ref{eq:bagnold}) and $l_s\approx 0.1$ was taken constant.}
\label{fig:v}
\end{center}
\end{figure} \noindent

For a qualitative estimate of the effects of Eq.(\ref{eq:wind}) onto
the sand transport over a dune, it is useful to consider once again
the local zeroth order model for aeolian sand transport,
Eq.(\ref{eq:q0}), i.e.\ a completely saturated flux $q=q_s(\tau)$ with
$q_s$ given by Eq.(\ref{eq:bagnold}). (Below, we will show that this
is asymptotically valid on large dunes in strong winds.) The distinct
features of Eq.(\ref{eq:wind}) that are missing in the zeroth order
approximation for the shear stress, Eq.(\ref{eq:tau0}), are then
easily seen to have potentially profound effects on the shape
evolution.  First, due to the depression of the shear stress at the
tails of the profiles, deposition rather than erosion may occur at the
windward foot. Secondly, due to the asymmetric contribution in
$\tau(x)$ there can be a net deposition on a symmetric heap of sand.
In particular, the shift of the position of the maximum shear stress
with respect to the top of the heap allows deposition at the top of
the heap. For an initially flat heap of sand there is thus the
possibility of a steepening of the windward slope and mass
growth. This implies that a plane sand surface is unstable against
modulations. To illustrate this effect, we used
Eq.(\ref{eq:solutions}) to calculate the migration velocity $v(x)$ of
a cosine--shaped heap of sand $f^2_C(x)$ according to
Eqs.(\ref{eq:v_def}), (\ref{eq:bagnold}), (\ref{eq:q0}), and
(\ref{eq:wind}). The latter is shown in the upper part of
Fig.~\ref{fig:v}.  The decrease of $v(x)$ on the lee side reveals the
anticipated self--amplifying tendency of the unstable lee slope to
steepen, since $v'<0$ implies that $v$ increases with height. This
gives rise to the formation of the slip face.  More interestingly,
Eq.(\ref{eq:wind}) renders $v(x)$ approximately constant over almost
the whole windward side if $\varepsilon=H\!/\!L$ is close to a certain
value determined by the values of the coefficients $A$ and $B$ in
Eq.(\ref{eq:wind}).  Slightly better constancy can be achieved for
slightly lower $n$ (with slightly larger $\varepsilon$) but \emph{not}
for the profiles $f_G$ and $f_L$, for which $v(x)$ is always
non--uniform.  Let us finally consider the dashed and dotted lines in
the upper part of Fig.~\ref{fig:v}. They were obtained for a smaller
and a larger aspect ratio and represent a steepening ($v'<0$) and
flattening ($v'>0$) of the windward side, respectively. In other
words, profiles with larger/smaller windward slopes are driven towards
the solution with constant windward $v(x)$ and a stable optimum
windward slope different from zero.  Altogether, Fig.~\ref{fig:v} thus
suggests that the coupled Eqs.(\ref{eq:bagnold}) and (\ref{eq:wind})
drive a heap of sand towards a ``dune'' with a cosine--like windward
profile of a preferred aspect ratio, and a slip face on the lee side.

From the qualitative analysis presented so far, it is not yet obvious
that this process converges to a translation invariant steady--state
solution.  Several previous studies using similar descriptions either
did not scrutinize the long time behavior of their models
\cite{Wippermann86,van_dijk-arens-boxel:99}, or failed to obtain
stable dunes \cite{zeman-jensen:88,Stam97}.  To obtain a consistent
general model for dune formation under general influx and wind
conditions, the present wind model Eqs.(\ref{eq:hunt}),
(\ref{eq:wind}) has to be appropriately adapted to situations
with flow separation above slip faces. And, most importantly, the
saturated--flux approximation Eq.(\ref{eq:q0}) of the ``zeroth order''
model has to be abandoned. These steps will be discussed in the
following subsection and in Section \ref{sec:saltation}, where we will
also explain the lower part of Fig~\ref{fig:v}. This will complete the
definition of the minimal model.  Its numerical solutions will be
presented in Section \ref{sec:solution}.

\subsection{Flow separation}\label{sec:separation}
\begin{figure}[t]
  \psfrag{s}[b]{streamlines} 
  \psfrag{sb}[cb]{separation bubble}
  \psfrag{ss}[cb]{separating streamline}
  \psfrag{d}{dune}
\begin{center}
  \includegraphics[width=\columnwidth]{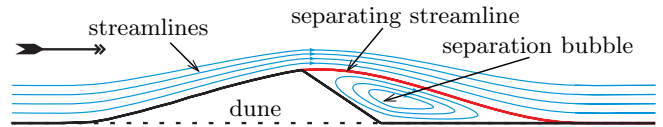}
  \caption{Sketch of the central slice of a barchan dune and the
  separation bubble. The shear stress on the windward side of the dune
  is calculated by applying Eq.(\ref{eq:wind}) to the
  phenomenologically defined envelope of the dune and the separation
  zone.}
\label{fig:separation}
\end{center}
\end{figure}

\begin{figure}[tb]
  \begin{center}
   
    \includegraphics[width=\columnwidth]{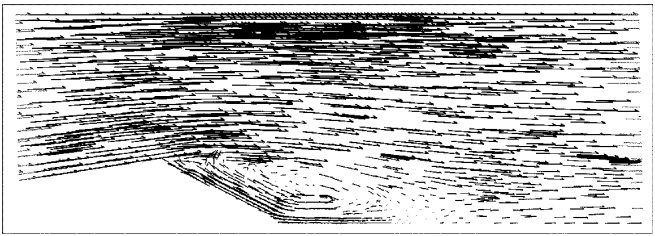}
    \caption{Cut along the symmetry plane of a $3d$ barchan dune. The
      velocity vectors calculated numerically with a commercial 
      fluid dynamics solver \cite{Fluent5} clearly display the flow 
      separation at the brink and a large eddy in the wake region.}
      \label{fig:fluent}
\end{center}
\end{figure}

The wind model as discussed so far works fine for smooth heaps with
gentle slopes. However, as we have already mentioned, its application
to dune profiles with slip faces and sharp brink lines is not
straightforward. The perturbative turbulent boundary layer approach
leading to Eq.(\ref{eq:hunt}) does not account for flow separation, a
phenomenon that occurs at sharp edges and steep slopes to prevent an
extreme bending of the streamlines \cite{landau-lifshitz:fm}. (For
some of the technical terms involved in this section, the reader is
referred to Fig.~\ref{fig:separation}.)  Instead of bending the
streamlines around sharp edges, re--circulating eddies separate from
the (on average) laminar main flow, thereby creating an effective
envelope that diverts the main flow on a smooth detour around the
obstacle. See Figs.~\ref{fig:separation} and \ref{fig:fluent} for a
schematic sketch and a numerical calculation of a typical velocity
field, respectively. Fortunately, it turns out that dune formation and
migration do not in general depend very sensitively on the details of
this complicated process. Or in other words, there is a large number
of interesting problems of aeolian sand transport for which these
details are largely irrelevant, and for which their somewhat realistic
physical representation would create a huge overhead in complexity
(especially in $3d$) to an otherwise tractable problem. It was
therefore suggested earlier \cite{zeman-jensen:88} that for the
purpose of calculating the shear stress on the windward side of a
dune, one may to a good approximation represent flow separation on the
lee side by the following heuristic method. A wind model such as
Eq.(\ref{eq:wind}) restricted to smooth, gently sloped objects is
applied to the envelope
\begin{equation}\label{eq:envelope}
\tilde h(x)=\text{max} \{h(x),s(x)\}
\end{equation}
of a dune $h(x)$ and a phenomenologically defined separation bubble
$s(x)$. This disregards the fact that the separating streamline does
not represent a solid boundary of the same roughness as the original
object, but the corresponding errors are expected to be small.
Typically, one wants $s(x)$ to be a mathematically simple smooth
continuation of the dune profile. It is, however, crucial that the
latter respects some major phenomenological properties of flow
separation \cite{badbubble}.  Although this is by no means a rigorous
procedure, one can test its predictions for selected cases against
numerical solutions of various turbulence models. The hope is that via
this approach, one can eventually get a qualitative understanding of
the mechanisms and phenomena involved in dune formation and migration,
leaving certain quantitative aspects to a more elaborate (and much
more laborious) future analysis.

\begin{figure}
     \psfrag{X}[t]{$(x-x_d)/h_d$}
     \psfrag{T}{$\frac{\displaystyle s(x)}{\displaystyle h_d}$}
 \begin{center}
    \includegraphics[width=\columnwidth]{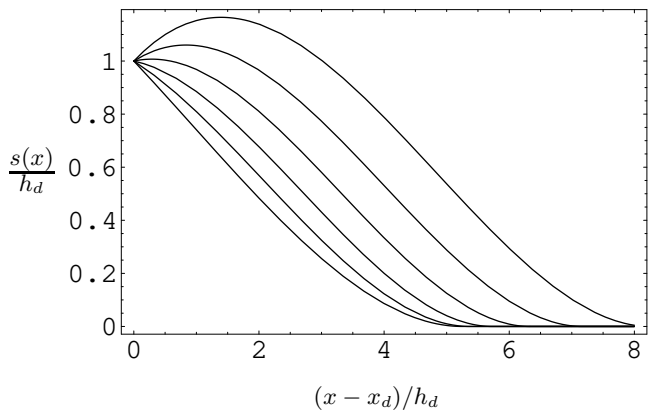} 
    \caption{Separation bubbles with a maximum negative slope of 0.25 
    according to Eq.(\ref{eq:sx}) for varying initial slopes $-0.25 \leq 
    h_d' \leq 0.25$. (The aspect ratio of the plot was stretched for 
    presentation.)}
      \label{fig:bubbles} 
\end{center}
\end{figure}

\begin{figure}
    \psfrag{x in L}[t]{$x / L$}
    \psfrag{h in h0, tau in tau0}{$h / H$, $\tau / \tau_0$}
    \psfrag{hxxxxxx}[lc]{\footnotesize $h$}
    \psfrag{sxxxxxx}[lc]{\footnotesize $s$}
    \psfrag{tauxxxx}[lc]{\footnotesize $\tau$}
 \begin{center}   
    \includegraphics[width=\columnwidth]{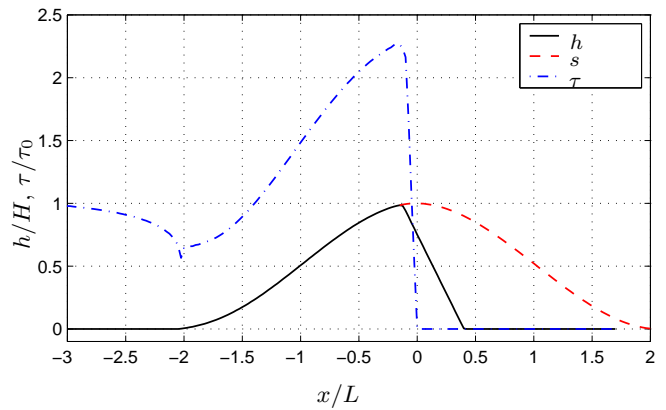}
    \caption{The windward profile $h(x)$ of a dune with slip face and
the separation bubble $s(x)$ form together a smooth effective
obstacle, defined by the envelope $\tilde h (x)$. To calculate the shear stress
$\tau(x)$ on the windward side of the dune, $\tilde h$ is substituted
for $h$ in Eq.(\ref{eq:wind}). In the region of re--circulation
the surface shear stress $\tau$ is set to zero \cite{rainout}. 
Without the separation bubble, $\tau(x)$ would develop a sharp
singularity at the brink.}
\label{fig:tau_bubble}
\end{center}
\end{figure}

In the spirit of the minimal model we want to parameterize the
separating streamline $s(x)$ in the simplest form that obeys
physically motivated boundary conditions at its detachment and
reattachment points $x_d$ and $x_r$. At detachment, the slope of the
separating streamline must match the slope of the dune. Moreover, also
the curvature must be continuous there, since discontinuities in
curvature are detected by Eq.(\ref{eq:wind}) and cause kinks in $\tau$
and discontinuous steps in the erosion/deposition as is e.g.\ the case
for the profile $f^2_C$. For the reattachment point, there are no
comparable restrictions to the slope and curvature, since the
separation bubble is not sharply defined there, and the model aims at
a realistic description of the conditions in the wake region only
insofar as they affect the shear stress on the \emph{windward}
side. On the lee side, inside the separation bubble, the shear stress
can simply be set to zero \cite{rainout}, since it is typically below
the threshold for aeolian sand transport.  Therefore, the choice of
the reattachment matching condition is a matter of convenience rather
than physical significance in the present model. However, we want
$s(x)$ to reproduce some common phenomenological knowledge about flow
separation. First, from many numerical calculations it is known that,
at high Reynolds numbers, the turbulent boundary layer reattaches at a
distance of about $6H$ after a backward--facing step of height
$H$. Secondly, it has often been observed experimentally that in
strongly turbulent flows over hills and symmetric triangular
obstacles, flow separation sets in if the backward slope exceeds an
angle of about $14^\circ$. Although, in both cases the exact numerical
values depend on various factors such as the surface roughness and the
Reynolds number, they shall be treated as fixed phenomenological
constants at the present stage. A model that fulfills all the above
requirements is a third order polynomial with continuous slopes at the
boundaries and a maximum negative slope of $\tan 14^\circ$. The
boundary conditions
\begin{equation}\label{eq:bc}
\begin{split}
s(x_d) & = h_d \equiv h(x_d) \qquad \; s(x_r) = 0 \\
s'(x_d) & = h_d' \equiv h'(x_d) \qquad s'(x_r) = 0 \;. \\
\bar s' & \equiv \text{max}\{-s'(x)\} = \tan 14^\circ = 0.25
\end{split}
\end{equation}
constrain the third order bubble parameterization to be of the form
\begin{equation}\label{eq:sepbubble}
  s(z)= (2 h_d + h_d' L_b)z^3 -(3 h_d + 2 h_d' L_b) z^2 + h'_d L_bz +
  h_d
\end{equation}
with $z\equiv (x-x_d)/L_b \in [0,1]$.  With the further abbreviation
$\nu\equiv h_d'/\bar s'$, we can express the length $L_b\equiv
x_r-x_d $ of the bubble as
\begin{equation}\label{eq:sx}
L_b =\frac{3h_d}{h_d'} \frac{1 - \nu - \sqrt{1 +  \nu}}{3- \nu} \approx
  \frac{3h_d}{2\bar s'}\left(1+ \frac  \nu4 + \frac{ \nu^2}8\right) \;,
\end{equation}
where the final approximation for small $h_d'$ is sufficient for our
purpose (and numerically better behaved as the exact expression). A
subtlety of such a separation bubble parameterization is the fact that
the slope at $x_d$ determines the length of the bubble, which in turn,
via Eq.(\ref{eq:wind}) influences the curvature at $x_d$. In other
words, the presence of the bubble introduces a non--local feed--back
between the slope and the curvature at the brink, which we believe is
physically reasonable. In Fig.~\ref{fig:bubbles} we give some examples
of separation bubbles for different boundary slopes $h_d'$, while
Fig.~\ref{fig:tau_bubble} illustrates the application of the above
discussion for the calculation of the shear stress.  It shows an
example for a dune profile $h(x)$ with slip face and the separation
bubble $s(x)$, together with the shear stress $\tau(x)$ resulting from
Eq.(\ref{eq:wind}) if $h$ is replaced by the envelope $\tilde h$.

We have performed several series of numerical fluid dynamics
calculations in $2d$ and $3d$ with the commercial fluid dynamics
solver Fluent~5 \cite{Fluent5} using the $k\epsilon$ and
\emph{large--eddy} turbulent closure models to confirm the general
picture outlined above and our particular implementation of the
separation bubble. The differences between numerical and theoretical
predictions for the shear stress on the windward side of various
dune-- and heap--like objects in $2d$ and $3d$ were quantitatively
small and not more significant than other neglected terms. Moreover, a
comparison of predictions obtained from Eq.(\ref{eq:sepbubble}) with
wind measurements on a barchan dune in Brazil \cite{sauermann:phd}
showed good agreement. Therefore, we are confident that the proposed
mathematical description of the wind shear stress captures the
relevant aspects in the spirit of the minimal model. As an example for
the numerical fluid dynamics calculation, we show in
Fig.~\ref{fig:fluent} the flow velocity in the symmetry plane of a
$3d$ barchan dune obtained with Fluent~5 \cite{Fluent5}. The wind is
blowing from left to right. The boundaries were chosen to be periodic
in the transverse direction. At the influx boundary, the velocity was
fixed by imposing a logarithmic velocity profile. The wind profile at
the outflux boundary is not known \emph{a priori}. Although, for high
Reynolds numbers the latter is expected to affect the solution only
close to the boundary, it is well known that different choices for the
outflux boundary condition as well as different discretization schemes
may lead to quantitatively different results \cite{wesseling:2000}.
Here, we chose to set the derivative of the velocity normal to the
outflux boundary to zero. The surface profile was represented as a
solid boundary with constant roughness length. Finally, along the top
boundary we imposed the velocity of the undisturbed logarithmic inflow
profile. The whole calculation was performed on a grid that had an
exponentially growing mesh size in the vertical direction. A
considerable grid refinement was necessary in the wedge--like region
of the separation bubble close to the brink.
 
These remarks complete the first task of constructing a model for the
calculation of the wind shear stress on a given dune profile as
outlined in Eq.(\ref{eq:task1}). By deriving the linear
Eq.(\ref{eq:wind}) for the shear stress and combining it with
the heuristic separation bubble, we have obtained an approximate but
numerically extremely efficient model for the wind shear stress on
dunes. This is a crucial step in the construction of a minimal model
of aeolian sand dunes, since the enormous complexity of the turbulent
air flow over structured terrain otherwise severely restricts the
possible applications of the model.

Going back to the upper part of Fig.~\ref{fig:v} with the above
discussion in mind, we can re--interpret this figure in order to
anticipate the behavior of the surface migration velocity $v(x)$ of a
dune with slip face. If, for qualitative purposes, $f^2_C$ is
interpreted as the envelope of a dune and its separation bubble, we
can conclude that the slip face must be located near the sharp drop of
$v(x)$ slightly upwind from the top of the envelope. This is indeed
consistent with observations for large dunes. Together with the good
representation of the windward profiles of large dunes
\cite{sauermann-etal:2000} by $f^n_C$ ($n\approx 2$), it suggests that
the given description becomes qualitatively correct in the limit of
large dune sizes.  The next section is devoted to the discussion of
important subtleties related to the fact that dunes are not typically
in this limit.

\section{Sand flux}\label{sec:saltation}
As outlined in Eq.(\ref{eq:task2}), the second task in the
specification of the minimal model is to find a prescription for
calculating the sand flux $q(x)$ for a given topography $h(x)$ and
shear stress $\tau(x)$. So far, we have been using the local
saturated--flux approximation Eq.(\ref{eq:q0}) in our qualitative
arguments. However, a closer look at the predictions obtained within
this approximation reveals a number of inconsistencies.  First, as we
have already noted in the discussion of Fig.~\ref{fig:v}, the use of
Eq.(\ref{eq:q0}) together with the complete wind model of
Section~\ref{sec:wind} leads to the odd prediction of deposition at
the windward foot of an isolated heap or dune, where the shear stress
decreases. This defect of Eq.(\ref{eq:q0}) has been noticed in the
literature before (see e.g.\ Refs.\
\cite{Wiggs96,sauermann-kroy-herrmann:2001}). Previous numerical
studies tried to avoid this problem by focusing onto the short time
behavior and by introducing \emph{ad hoc} heuristic methods such as a
``smoothing operator'' \cite{Wippermann86} or an ``adaptation length''
\cite{van_dijk-arens-boxel:99}. The reason for the problem is that the
saturated--flux approximation breaks down at a boundary
ground/sand. As another shortcoming, we want to mention that the model
as discussed so far predicts a universal scale invariant dune shape
with a brink that is displaced slightly upwind from the maximum of the
envelope, leading always to a positive slope at the brink.  A glance
at a real dune field proves that the latter is not always the case and
careful measurements \cite{sauermann-etal:2000} have revealed
systematic deviations from scale invariance.  Though less obvious, it
turns out that the reason for this discrepancy lies again in the
saturated--flux approximation. Both mentioned problems are thus
naturally resolved by introducing a slightly more general sand
transport law that allows for saturation transients.

\subsection{Saturation transients}
The saturated--flux approximation Eq.(\ref{eq:q0}) assumes that the
flux is everywhere equal to the equilibrium transport capacity $q_s$
of the wind. However, due to variable wind or sand conditions, the
actual sand flux $q$ is in general different from $q_s$. These
deviations are called saturation transients, because they quickly
relax to zero under homogeneous conditions. We have recently
demonstrated \cite{sauermann-kroy-herrmann:2001} that this relaxation
occurs within a characteristic length scale, called the
\emph{saturation length} $\ell_s$, which is related to (but distinct
from) the mean saltation length of the grains. It was moreover shown
how the introduction of saturation transients cures the problem of
deposition at the windward foot of an isolated sand dune.  Here, we
only summarize the most pertinent results of this earlier development
in order to demonstrate how a size dependence of the dune shape
naturally results as a consequence of saturation transients.

The sand transport model of Ref~\cite{sauermann-kroy-herrmann:2001} is
based on a mean--field like description of saltation. It models a
typical grain that is accelerated by friction with the air and slowed
down by dissipative interactions with the bed. The average properties
of the complicated splash process
\cite{Anderson91,Nalpanis93,Rioual2000} are subsumed into two
dimensionless parameters, an effective restitution coefficient
$\alpha$ for collisions with the bed, and a kinetic coefficient
$\gamma$ that characterizes the relaxation of the density of saltating
grains to its saturated value. Together with an effective height for
the wind--grain interaction that enters only logarithmically, these
are the only phenomenological parameters of the model. They have been
determined by a comparison with experiments and grain scale
simulations.  Formally, the model consists of two coupled differential
equations for mass and momentum conservation, and a modified turbulent
closure relation that accounts for the feedback of the saltating
grains on the wind velocity.

\begin{figure}[tb]
 \psfrag{T}{$\tau/\tau_t$}
 \psfrag{L}{$\ell_s \;\text{[m]}$}
  \begin{center}
    \includegraphics[width=\columnwidth]{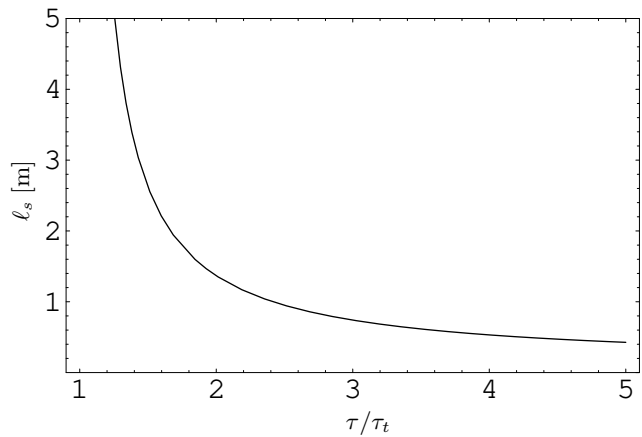} 
    \caption{The
    saturation length $\ell_s$ in meters as a function of the shear
    stress exerted by the wind onto the sand bed. This function 
    sets the natural length scale for dunes and heaps.}
\label{fig:ls}
\end{center}
\end{figure}

For the present purpose, the model can be simplified by taking
advantage of the fact that the prevailing conditions in applications
to dunes are typically well described by the steady--state
($\partial/\partial t\simeq 0$) version. Further, the relaxation of
the typical sand transport velocity can be assumed to be fast compared to
the variations in the density of mobilized grains in the saltation
layer. Approximating the latter by its saturated value for the
calculation of the effective wind speed $u_{\rm eff}$ via the modified
turbulent closure, one can decouple the mass and momentum conservation
equations.  The whole model can then in a reasonable approximation be
reduced to a single differential equation
\begin{equation}
  \label{eq:flux}
  \ell_s\partial q/\partial x = q( 1-q/q_s)  
\end{equation}
for the sand flux $q(x)$.  The shear stress dependent parameters
\begin{equation}\label{eq:ls}
  \ell_s= l/(\tau/\tau_t-1)\;, \qquad q_s = \rho_s u_s 
\end{equation}
are immediately identified as the saturation length and the saturated
flux, respectively.  The equation for $q_s$ generalizes
Eq.(\ref{eq:bagnold}) to arbitrary wind speeds. In the following we
specify the explicit expressions for both quantities as they result
from the sand transport model of
Ref.~\cite{sauermann-kroy-herrmann:2001}, but the structure of
Eq.(\ref{eq:flux}) is thought to be more general and independent of
the precise form of Eq.(\ref{eq:ls}). Again, $\tau(x)$ is the position
dependent shear stress discussed in Section~\ref{sec:wind} and $\tau_t
\approx 0.1\; \text{kg\,m}^{-1}\text{s}^{-1}$ is the estimated impact
shear stress threshold that corresponds to a critical shear velocity
$u_{*t}\approx 0.28 \;\text{m\,s}^{-1}$ \cite{Owen64}.  (For
simplicity, we do not introduce the additional threshold for purely
aerodynamic entrainment here, but allow instead for a small residual
influx even if the latter is nominally zero.) To make the underlying
structure of the model more palpable, we have expressed $\ell_s$ in
terms of another characteristic length scale $l\equiv 2\alpha
u_s^2/(g\gamma)$, which (up to a numerical factor) is the average
saltation length of the grains. The latter --- but not $\ell_s$ ---
must always be considerably smaller than the dune length for the model
to be applicable.  Further, we have decomposed $q_s$ into the
saturated density $\rho_s = 2\alpha(\tau-\tau_t)/g$ and the effective
sand transport velocity at saturation $u_s = u_{\rm eff} - \delta u$
with $u_{\rm eff}$ the effective wind velocity that accelerates the
grains, given by
\begin{equation}\label{eq:ueff}
 u_{\rm eff}\kappa\sqrt{\varrho_{a}} = 2\sqrt{\tau_t+(\tau
-\tau_t)/\zeta} + (\ln\zeta'-2)\sqrt{\tau_t}\;. 
\end{equation}
By $g$ we have denoted the gravitational acceleration and from
Ref~\cite{sauermann-kroy-herrmann:2001} we adopt the (approximate)
numerical values $\alpha=0.35$, $\gamma=0.2$, $\zeta= 8$, $\zeta'=
200$, and $\delta u= 1.8\;$m/s for the lag velocity of the grains. We
note that these numerical values are not completely independent of
each other and of the mentioned value for the impact threshold
$\tau_t$, due to the calibration of the sand transport model with
experimental data \cite{sauermann-kroy-herrmann:2001}.  For
convenience we show a plot of the saturation length $\ell_s$ obtained
with these values in Fig.~\ref{fig:ls}.  This completes the definition
of the sand transport part of the minimal model on gently sloped
ground.

\subsection{Consequences}
Before we complete the general model definition by a brief paragraph
on slip faces, we want to point out some implications of the model as
developed so far.  First, note that the full expression for $q_s$
given in Eq.(\ref{eq:ls}), contains Eq.(\ref{eq:bagnold}) as a
limiting case for strong winds but is better approximated by
$q_s\propto \tau-\tau_t$ for moderate wind speeds. For weak flux
gradients and strong winds, one may set the left hand side of
Eq.(\ref{eq:flux}) to zero, leading to $q=q_s$. This is
typically the case on most of the windward slope of a large dune,
where the left hand side of Eq.(\ref{eq:flux}) can roughly be
estimated by $\ell_s q_s/L \ll q_s$.  The local saturated--flux
approximation with Eq.(\ref{eq:bagnold}) for $q_s$, which we have
applied throughout our qualitative discussion so far, is thus
asymptotically valid for large dunes and strong winds (except near the
windward foot of an isolated dune). This is what one might have
expected in the first place, and the reader may wonder at this point
how the saturation transients and their characteristic length scale
$\ell_s$ can have the claimed importance. How can $\ell_s$ affect the
shape of a dune that is typically about two orders of magnitude
larger? This apparent puzzle is now easily resolved by going back to
Figs.~\ref{fig:hills}, \ref{fig:sym_asym} and Eq.(\ref{eq:shift}) and
by observing that the symmetry breaking shift $\delta x_\tau$ of the
location of the maximum of the shear stress with respect to that of
the maximum of the height profile (or envelope) that is responsible
for the finite windward slope and growth of dunes, is also of the
order of some per cent of the total dune length. In summary, the
longitudinal profile of dunes and heaps is determined by the
competition of two quantitatively small but qualitatively crucial
effects, one related to turbulent wind flow and the other to sediment
transport. This may be the reason, why its explanation proved elusive
for a long time.

To get a qualitative idea of the consequences of the introduction of
the generalized non--local flux law in Eq.(\ref{eq:flux}) as
replacement for Eq.(\ref{eq:q0}), we want again to go back to our
discussion of the surface migration velocity of the cosine shaped heap
$f^n_C$ in Fig.~\ref{fig:v}. Let us for the moment adopt a crude
approximation and replace the expression for $q_s$ given in
Eq.(\ref{eq:ls}) by its simpler limiting form $q_s \propto \tau^{3/2}$
introduced in Eq.(\ref{eq:bagnold}). We also neglect the variation of
$\ell_s$ on the dune and replace it by a (fine--tuned) constant
$\ell_s\approx 0.1\, L$.  With an influx (about $0.7\, q_s$ for the
solid and $0.8\, q_s$ for the dashed line in the lower part of
Fig.~\ref{fig:v}) one can thus achieve a fairly constant surface
velocity over the \emph{whole} length of a cosine shaped heap.  Again
the constancy is slightly better for $n<2$ than for $n=2$. It is
further improved by reducing the slope of the heap well below the
optimum windward slope of the dune obtained for $\ell_s\to0$, as seen
from a comparison of the solid line and the dashed line. We also note
in passing that the influx needed to maintain the shape is increasing
with decreasing slope. Together, the plots in Fig.~\ref{fig:v}
confirms our claim that even an $\ell_s\ll L$ may visibly affect the
overall shape of aeolian dunes.  Although, for the example shown in
the lower part of Fig.~\ref{fig:v}, the saturation length is only
about $1/30$ of the heap length, the slip face instability is
evidently completely washed out. Altogether, this strongly suggests
the existence of translation invariant cosine shaped heap solutions
for the model.  The ultimate proof will be provided by the numerical
results presented in Section~\ref{sec:solution}, where the full form
of $q_s$ and $\ell_s$ according to Eq.(\ref{eq:ls}) will be used, but
the present crude approximation already illustrates the main point,
and also demonstrates that the behavior is a generic consequence of
Eq.(\ref{eq:ls}) and insensitive to the detailed form of the
parameters $\ell_s(x)$ and $q_s(x)$ that may phenomenologically be
somewhat different from the model prediction without affecting our
general conclusions.

An immediate consequence of the foregoing discussion is the existence
of a minimum dune size. For small enough dunes, the slip face
instability is washed out by the saturation transients. One may also
arrive at this conclusion from an analysis of heaps. To this end we
observe that the value of the saturation length $\ell_s$ is a property
of the wind velocity and the saltation kinetics and depends on the
topography only indirectly through the variable shear stress
$\tau$. Moreover, it is apparent from Fig.~\ref{fig:ls} that this
dependence becomes weak for strong winds. On the other hand, the
symmetry breaking shift $\delta x_\tau$ is proportional to the
absolute size of the heap (or envelope) and not directly dependent on
the wind velocity. For the special profile $f_L$, this was verified
analytically in Eq.(\ref{eq:shift}). As we have seen, a smooth heap
can only be a translation invariant solution of the model if the lag
(of order $\ell_s$) of $q(x)$ with respect to $q_s(x)$ and the shift
$\delta x_\tau$ are fine--tuned to guarantee a vanishing erosion rate
at the top of the heap. From this we expect heaps to obey
\begin{equation}\label{eq:heap_aspect}
\ell_s \simeq \delta x_\tau \propto L \approx \text{const.}
\end{equation}
to a first approximation. This condition can only be fulfilled if the
aspect ratio $\varepsilon$ of heaps grows proportional to their size
(i.e.\ roughly $\varepsilon\propto H$). Hence, in contrast to large
dunes with slip face, for which we have argued that they are
asymptotically scale invariant ($\varepsilon \sim$ const.), heaps must
have a strongly size dependent aspect ratio. As a consequence,
translation invariant heap solutions obviously cannot exist beyond a
certain critical size. A slip face will develop when the shear stress
on the lee side of the heap drops below the threshold value $\tau_t$,
or at the latest, when the lee slope exceeds the critical slope for
flow separation. This will be further analyzed in
Section~\ref{sec:solution}.  Finally, we note that the steady--state
flux of a heap can be estimated by the observation that the outflux is
essentially determined by the strength of the reduction
$\tau_0-\tau_{\rm min}$ of the shear stress at the lee end of the
heap. According to Eq.(\ref{eq:scaling}), the latter is (for a given
shape) proportional to the aspect ratio $\varepsilon$. For qualitative
purposes, the outflux $q^{\rm out}$ may thus be estimated in the
saturated flux approximation with
Eqs.(\ref{eq:ls}),(\ref{eq:ueff}),(\ref{eq:scaling}) as
\begin{equation}\label{eq:outflux}
q_s^{\rm out} \propto \tau_{\rm min} -\tau_t \propto
\varepsilon_c-\varepsilon \;,
\end{equation}
where we have assumed $\tau_{\rm min}/\tau_t \lesssim 2$ (fulfilled
for moderate wind speeds and/or heaps near the critical heap size)
to linearize the expression for $q_s(\tau)$ given in
Eq.(\ref{eq:ls}). Here, $\varepsilon_c \propto \tau_0-\tau_t$ is the
critical aspect ratio for which the shear stress on the lee drops
below the threshold and the outflux vanishes. Note that the latter
increases with increasing shear stress whereas the heap length
decreases according to Eqs.(\ref{eq:ls}), (\ref{eq:heap_aspect}). The
effects of the two trends onto the critical heap mass could therefore
partially cancel unless the lee slope exceeds the critical slope for
flow separation.

\subsection{Slip face}\label{sec:slipface}
We have argued above that for large heaps ($L\gg\ell_s$), aeolian
sediment transport tends to increase the downwind slope until it
reaches the angle of repose of the grains. At this point, any further
increase of the lee slope initiates avalanches that restore a slope
slightly below the static angle of repose and eventually create a slip
face of a roughly uniform slope of about $32^\circ - 35^\circ$. Since
the physical modeling of this process itself is not a major objective
of the present contribution, we can choose between different possible
implementations for this phenomenon. In $2d$ it is possible to
represent the slip face as boundary condition for the sand transport.
It is uniquely determined by its fixed uniform slope and mass
conservation. However, with regard to a future generalization of the
present $2d$ model to $3d$ we chose a more realistic implementation
based on a widely--used avalanche model \cite{Bouchaud94}. The
formulation of this model bears some close similarities with the sand
transport model presented in the preceding paragraph, and thus
suggests itself as a natural extension of the latter to the slip
face. This completes the definition of the minimal model that will be
solved numerically in the next section.

\section{Solution of the model}\label{sec:solution}
Apart from the model definition, the preceding sections have provided
some qualitative insights into the main mechanisms responsible for
dune formation and migration. Now we are prepared to study numerically
the quantitative predictions of the model. Again, we emphasize that we
only can explore some major features of the model in the present
report, leaving many interesting questions and more systematic and
quantitative parameter studies for future work.

\begin{figure}[tb]
  \psfrag{initial}[c][c]{\tiny $h(x,t=0)$}
  \psfrag{bubble}[c][c]{\small Eq.(\ref{eq:sepbubble})}
  \psfrag{tau}[c][c]{\small Eq.(\ref{eq:fourier_wind})}
  \psfrag{flux}[c][c]{\small Eq.(\ref{eq:flux})}
  \psfrag{mass}[c][c]{$\genfrac{}{}{0pt}{1}{\text{Eq.(\ref{eq:mass})}\; \&}
    {\text{\tiny avalanches}}$}
  \begin{center}   
    \includegraphics[width=\columnwidth]{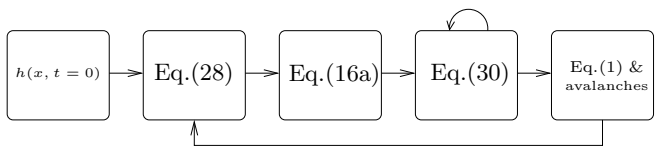}
    \caption{Solution of the minimal model.}
      \label{fig:flowchart} 
\end{center}
\end{figure}

For convenience, the solution procedure of the minimal model is
summarized as a flow chart in Fig.~\ref{fig:flowchart}. One starts
with an initial profile $h(x,t=0)$ (typically $f_G$ or $f_C^2$),
checks whether a separation bubble has to be added for the calculation
of the shear stress, then obtains the latter from
Eq.(\ref{eq:fourier_wind}) and uses the result as input for the
iterative solution of the sand transport equation
Eq.(\ref{eq:flux}). This finally gives the erosion/deposition needed
to update the surface profile. Technically, Eq.(\ref{eq:fourier_wind})
is implemented as a Fast--Fourier--Transform algorithm, and for the
integration of Eqs.(\ref{eq:flux}) and (\ref{eq:mass}) an upwind
discretization scheme is used. Simulation times can be reduced by
using an adaptive time step.

\subsection{Steady--state shapes}
The scheme of Fig.~\ref{fig:flowchart} can be iterated for different
influx boundary conditions. For all of the numerical calculations
presented below, we chose periodic boundary conditions. They are the
natural choice for studies of the steady--state shapes.  To
investigate the mass balance under prescribed influx conditions, on
the other hand, one has to apply open boundary conditions.

\begin{figure}[tb]
  \psfrag{h in m}[][90]{$h(x)$ [m]} 
  \psfrag{x in m}{$x$ [m]}
  \begin{center} 
\includegraphics[width=\columnwidth]{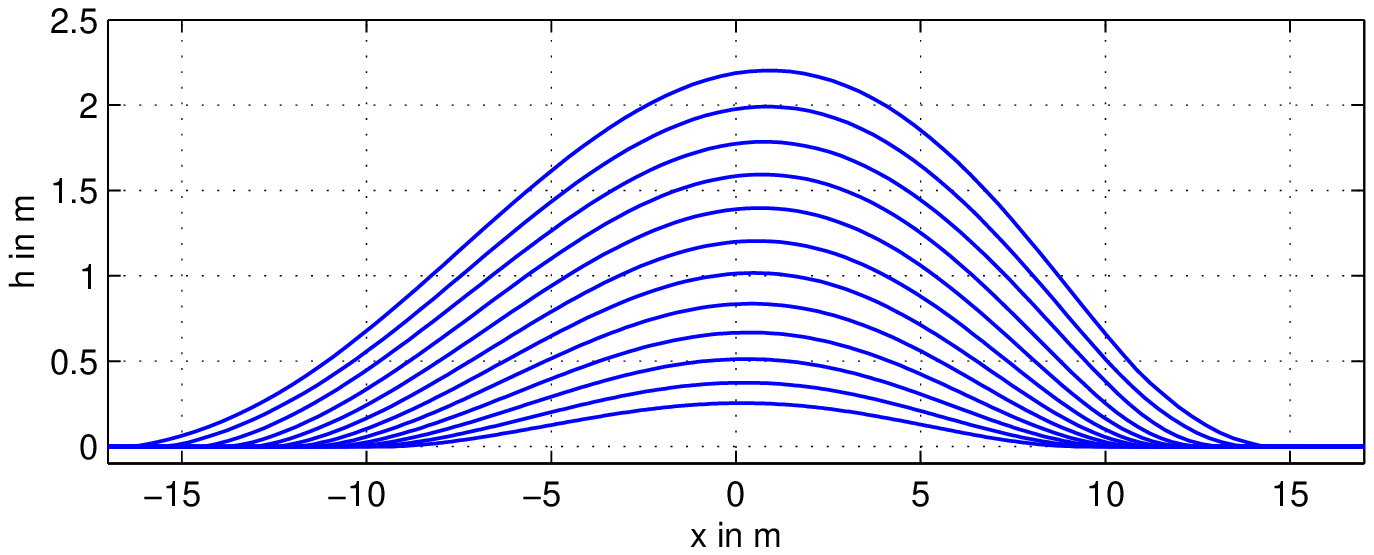}
\includegraphics[width=\columnwidth]{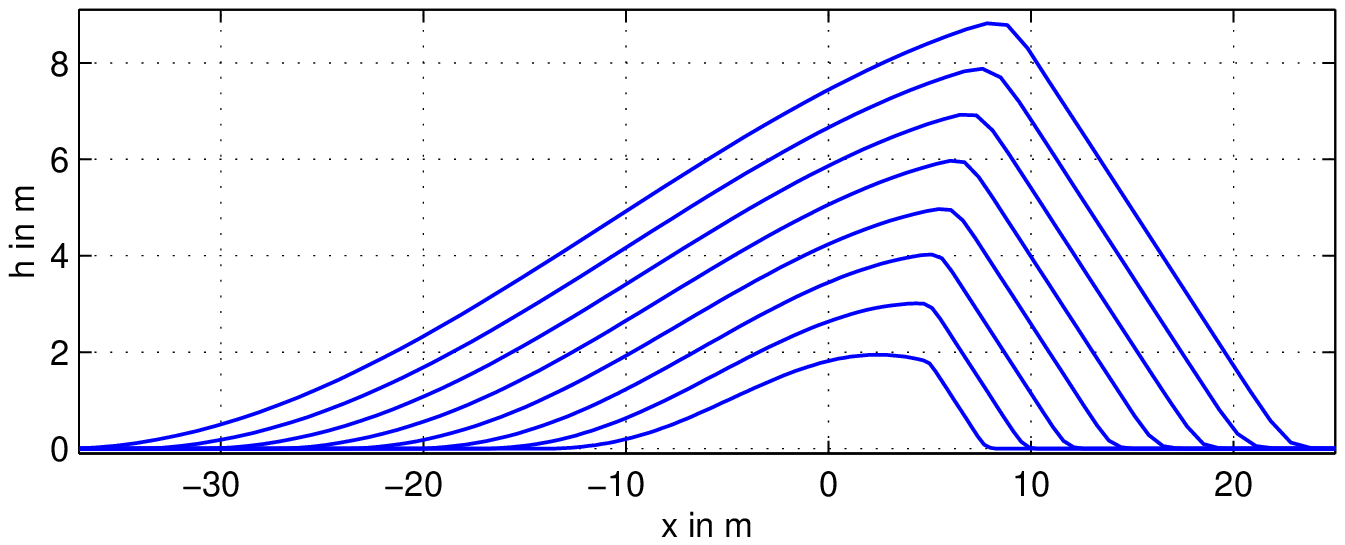}
  \caption{Steady--state heaps (upper plot) and dunes (lower
  plot). The aspect ratio is stretched for better visualization.}
  \label{fig:shapes}
\end{center}
\end{figure}

\begin{figure}[tb]
  \psfrag{flat}{\small flat}
  \psfrag{steep}{\small steep}
  \psfrag{H}[rl]{$H$ [m]}
  \psfrag{V}[t][b]{$H L$ [m$^2$]}
  \psfrag{infit}{\small $\propto H L$}
  \psfrag{asfit}{\small $\propto \sqrt{H L}$}
  \begin{center}   
    \includegraphics[width=\columnwidth]{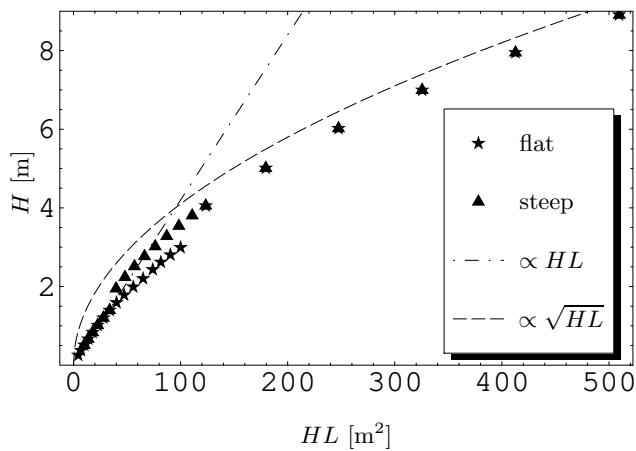}
   \caption{Steady--state heights $H$ versus the product of the height
  and length of the heaps and dunes. In the hysteretic
  regime, flat and steep initial conditions have to be distinguished.}
    \label{fig:H_V}
\end{center}
\end{figure}

Fig.~\ref{fig:shapes} shows steady--state solutions of the model for
initial profiles $f_G$ of different mass. These solutions are obtained
for fixed wind conditions with parameters $A=3.2$ and $B=0.25$
appropriate for the central slice of a $3d$ (symmetric) heap or of a
barchan dune. The shear velocity $u_*=0.4$ m/s lies well above the
impact threshold. (The situation very close to or below the threshold
would need special attention.) As anticipated above, large dunes
become asymptotically scale invariant. The asymptotic master curve for
the windward profile is practically indistinguishable from the profile
$f_C^n$ ($n\lesssim 2$), and the slope at the brink is indeed
positive. Its average windward slope is inversely proportional to the
value of the parameter $A$ given in Eq.(\ref{eq:a_b}). Due to the
additional terms in the expression Eq.(\ref{eq:correction}) for the
shear stress on dunes with a finite width, somewhat steeper average
windward slopes are predicted for barchan dunes than for transverse
dunes under identical influx and wind conditions. However, a detailed
quantitative comparison is probably beyond the scope of the present
semi--quantitative implementation. More important are the remarkable
qualitative predictions of the model. In particular, the fact that
dunes with slip face are only stable above a certain (wind dependent)
critical size, whereas smooth steady--state heaps only exist below a
critical size, deserves attention. We also note that the steady state
is not always unique. There is a hysteretic regime, where the initial
conditions can select one of two possible steady--state shapes and
accordingly the masses for the two sets of profiles in
Fig.~\ref{fig:shapes} are not all distinct. The largest heaps in the
upper plot were obtained from flat initial profiles $f_G$, whereas the
smallest dunes with slip face in the lower plot were obtained from
steeper initial profiles $f_G$ of the same mass. Especially, the dune
with a negative slope at the brink could only be obtained from steep
initial conditions. Since under natural wind and sand conditions, the
initial conditions themselves will generally be heaps or dunes close
to the steady state, one can say that the model predicts a critical
heap size for slip face formation and a critical dune size for slip
face destruction. In both cases the slip face is finite as a
consequence of flow separation.  The latter also allows a dune to be
somewhat higher than a heap of the same mass, since its effective
volume as seen by the average air flow is increased by the separation
zone. As anticipated, the aspect ratio of the dunes is asymptotically
constant, whereas it is strongly size--dependent for heaps. This
effect can be seen more quantitatively in the representation of
Fig.~\ref{fig:H_V}, where the height $H$ of the steady--state heaps
and dunes is plotted versus the product $HL$ of their height and
length $L$. Clearly, heaps are better described by $H\propto H L$ as
predicted in Eq.(\ref{eq:heap_aspect}), whereas large dunes approach
the scaling limit $H\propto \sqrt{H L}$.

\begin{figure}[tb]
  \begin{center} 

  \psfrag{ustar035xxxx}{\tiny $u_*=0.35$}
  \psfrag{ustar040xxxx}{\tiny $u_*=0.40$} 
  \psfrag{ustar045xxxx}{\tiny $u_*=0.45$} 
  \psfrag{ustar050xxxx}{\tiny $u_*=0.50$}
  \psfrag{propto1hxxxx}{\tiny $\propto 1/H$}
  \psfrag{propto1Lxxxx}{\tiny $\propto 1/L$} 
  \psfrag{height in m}[t][b]{$H$ [m]} 
  \psfrag{length in m}[t][b]{$L$ [m]} 
  \psfrag{vd in m per year}{$v$ [m/yr]} 
  \psfrag{Lvd in m per year}{}   
  \includegraphics[width=0.45\columnwidth]{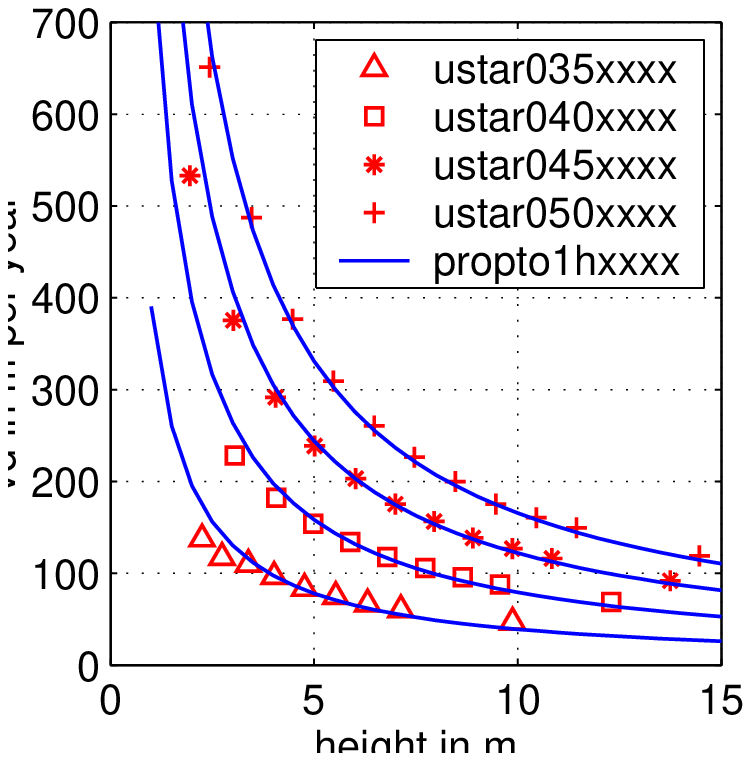}
  \includegraphics[width=0.45\columnwidth]{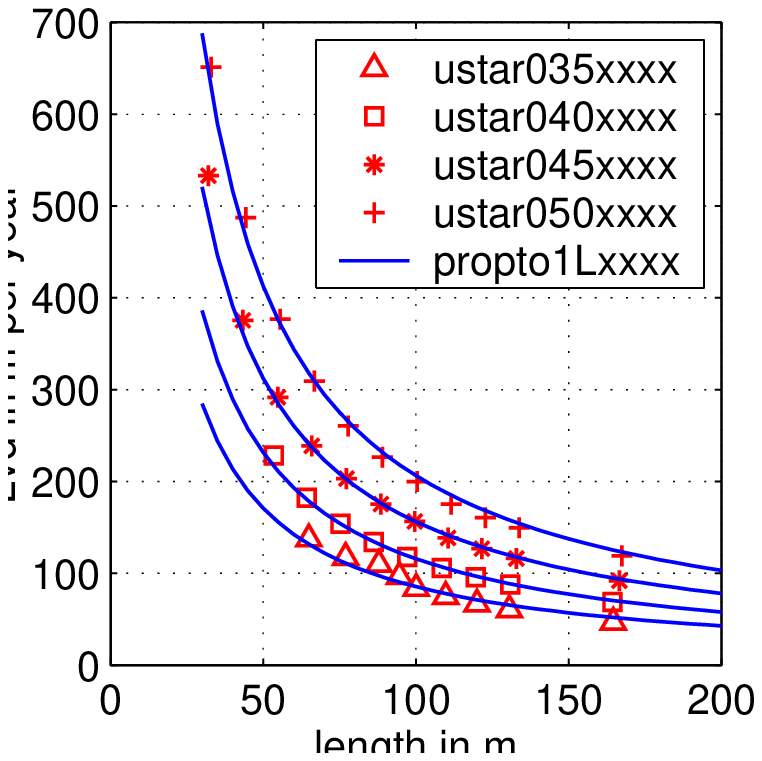}
  \caption{Migration velocities predicted by the minimal model for
  steady--state dunes of different size at various wind
  velocities. The caption gives the shear velocity $u_*$ in m/s. The
  numerical data are compared to the scaling laws $v\propto H^{-1}$
  (left) and $v\propto L^{-1}$, where $L$ is the length of the
  envelope of the dune and its separation bubble as described in the
  main text. (Note that the migration of real dunes is
  substantially slower due to the small fraction of wind days per
  year.)}  \label{fig:v_H_L}
\end{center}
\end{figure}

\subsection{Migration velocity}
For the overall migration velocity of steady--state dunes with a scale
invariant profile, we derived on general grounds the simple scaling
prediction $v\propto L^{-1}$ in Section~\ref{sec:v}. We have also
given some arguments why this prediction should be rather robust
against relaxing the condition of shape invariance, in contrast to the
relation $v\propto H^{-1}$ that can only be inferred from it if the
scaling assumption holds exactly. Here, we check these predictions
for the steady--state solutions numerically. Fig.~\ref{fig:v_H_L}
shows the numerically obtained migration velocity for dunes fitted to
the scaling relations $v\propto H^{-1}$ (left) and $v\propto L^{-1}$
(right).  As we have mentioned above, one has to take for $L$ the
characteristic length of the envelope rather than that of the dune
alone. For simplicity, we estimate $L$ by adding $6H$ to the
horizontal length from the windward foot of the dune to its crest,
thus neglecting the weak slope dependence of the size of the
separation bubble. Obviously, the $L^{-1}-$scaling is superior for
moderate winds and small dunes where $v\propto H^{-1}$ systematically
fails to describe the data. This is also supported by field data
\cite{Finkel59}. Both fits become identical in the scaling limit $L\gg
\ell_s$. Due to the decrease of $\ell_s$ with the wind speed, the
latter is reached for smaller dunes at stronger winds.

\subsection{Stability}
\begin{figure}[tb]
  \psfrag{flat}{ flat} 
  \psfrag{steep}{ steep} 
  \psfrag{qin}[c]{$q^{\rm out}$ [kg\,m$^{-1}$s$^{-1}$]}  
  \psfrag{eps}{$H/L$}
  \psfrag{growing}{$\stackrel{\displaystyle
  \text{growing}}{\text{solutions}}$}
  \psfrag{shrinking}{$\stackrel{\displaystyle
  \text{shrinking}}{\text{solutions}}$} 
\begin{center}
  \includegraphics[width=\columnwidth]{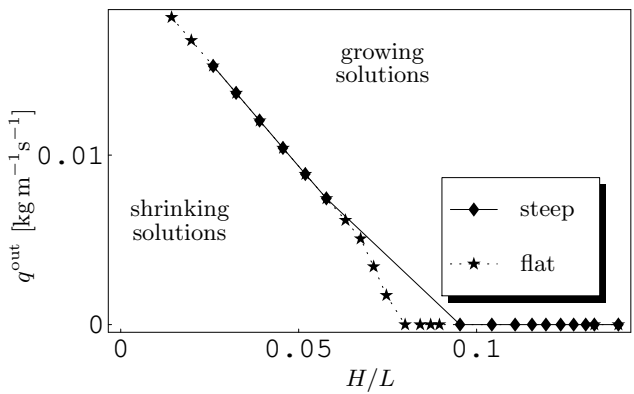}
  \caption{Steady--state outflux under periodic boundary
  conditions. In the hysteretic regime, steep and flat initial
  conditions have to be distinguished, as in Fig.~\ref{fig:H_V} The
  figure may also be read as a phase diagram for the situation with
  open boundary conditions. In this case the steady--state solutions
  --- though attractors for the shape --- are unstable with respect to
  mass fluctuations.}  \label{fig:q_V}
\end{center}
\end{figure}

We have already pointed out that the choice of different boundary
conditions for the flux allow a separate discussion of shape and mass
stability. This is of practical importance, since (in $2d$) all
steady--state shapes are unstable with respect to mass changes. If the
influx of a steady--state solution deviates slightly from its
corresponding steady--state flux, this solution will start to either
shrink until it has flattened out or grow without bound.  Though the
latter effect could (but need not) be a peculiarity of the vanishing
outflux for $2d$ dunes with slip face, at least the former generalizes
to $3d$ heaps. Despite the fact that the steady--state shapes are
(locally) stable attractors for the shape evolution under periodic
flux conditions, mass stability can in general not be achieved under
open boundary conditions. The situation is clarified in
Fig.~\ref{fig:q_V}. It depicts the steady--state sand flux $q$ over
the bedrock as a function of aspect ratio. The numerical results
nicely confirm our theoretical expectation from
Eq.(\ref{eq:outflux}). For all dunes with slip face the flux vanishes
identically in $2d$, whereas in general it grows with decreasing size
for smooth heaps.  For open boundary conditions, the line in
Fig.~\ref{fig:q_V} can be interpreted as an unstable phase boundary
(with hysteresis) between infinitely growing and shrinking
solutions. For example, a heap with influx slightly below the
steady--state, will shrink a bit. To remain close to the steady--state
shape, it will therefore mainly reduce its height, whereas its length
will stay almost constant. Due to the reduced aspect ratio
$\varepsilon$, $\hat \tau$ decreases in magnitude and the shear stress
depression at the lee boundary is less pronounced. As a consequence
there is less deposition on the downwind slope and the outflux is
higher, so that the heap shrinks even more etc. A completely analogous
reasoning applies to the opposite case of higher influx. The
corresponding shape attractors are the scale invariant asymptotic dune
shape and the flat surface, respectively.

The above discussion explains, why isolated smooth aeolian sand heaps
are rarely observed as distinct features of desert topographies. Under
approximately stationary wind and influx conditions they only exist as
transient states that either vanish or develop into dunes with slip
face. Under variable wind and influx conditions, the situation is less
clear and deserves a detailed study of its own. For example, the model
predicts that during a period of strong wind all dunes are driven
towards the asymptotic shape. After a subsequent period of weak winds,
finite size effects become more pronounced, and small dunes may
develop longitudinal profiles like those in the hysteretic regime or
even loose their slip face. Again, the case $\tau_0 \approx \tau_t$ of
a shear stress close to or below the threshold shear stress needs
special attention.

The prevailing wind conditions as well as recent changes in the wind
velocity are thus encoded in a complicated but comprehensible way in
the shapes of the dunes in a dune field. This is a promising direction
for further studies. One may hope that by systematic studies
along these lines one will in the future be able to infer flow
conditions in remote or uncomfortable places (e.g.\ on the sea bottom
or on other planets) by analyzing dune
shapes.

\subsection{Relaxation dynamics}\label{sec:relaxation}
\begin{figure}[t]
  \begin{center}  
    \includegraphics[width=\columnwidth]{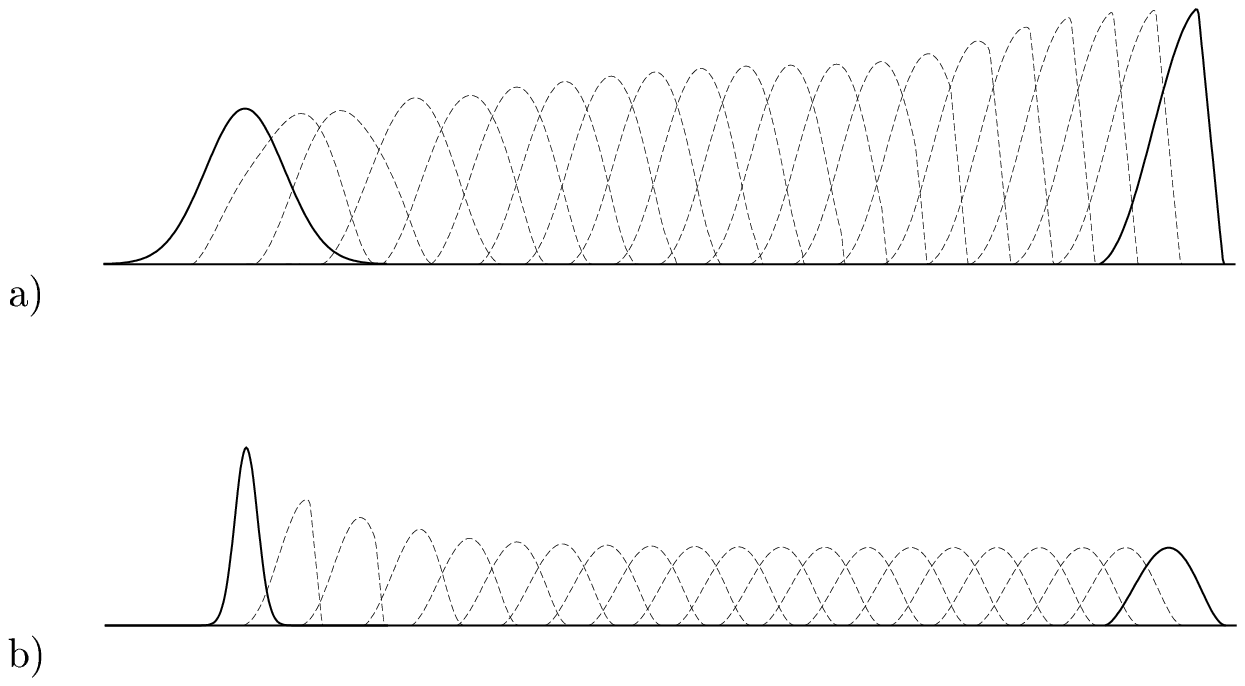}
    \caption{Growth histories for two Gaussian heaps of different mass
      and initial aspect ratio for periodic flux boundary conditions.
      Note that the distances to reach the steady--state shape are
      different.  (The aspect ratio of both plots was rescaled by a
      common factor for better visualization.)}
\label{fig:growth} 
\end{center}
\end{figure}

As a first step towards an understanding of the effects of variable
wind speeds (for constant wind direction), this section is devoted to
an exploratory study of the transient shape evolution. We restrict
ourselves to periodic boundary conditions leaving the richer phase
space of open boundary conditions for future
studies. Fig.~\ref{fig:growth} shows two extreme scenarios. A flat
initial condition with a mass greater than the critical mass for slip
face formation (a), and a steep initial condition with a mass below
the critical mass for slip face destruction (b). The steep initial
condition gives rise to the temporary formation of a slip face that is
finally washed out by the saturation transients, whereas the flat heap
steepens until the shear stress on the lee falls to the threshold
value.  This causes complete deposition on the lee side of the
heap. Whether this happens before or with the onset of flow separation
depends on wind and influx conditions. As it also depends on the
precise numerical values of some of the phenomenological parameters of
the model, a detailed parametric study is again beyond the scope of
the present contribution.

\noindent

Although the times to reach the steady state are apparently somewhat
longer for the flat initial condition, it is evident from
Fig.~\ref{fig:growth} that the relaxation dynamics is in general
relatively fast even if the initial condition is far from the steady
state shape. Large dunes under low influx conditions as they prevail
e.g.\ in fields of isolated dunes should therefore be well described
by an adiabatic approximation assuming that (except after drastic
changes in the wind and sand conditions as they occur during sand
storms) the dune is practically in a steady state.  Apart from the low
influx, this also relies on the fact that virtually no sand is lost
over the slip face. For a large isolated $3d$ barchan dune this
implies that most of the sediment transported over the dune is
actually trapped in a tread--milling flux, and only a small portion of
the total flux is contributed by and contributes to the external flux.
Hence, under steady wind conditions these dunes are in a quasi steady
state and thus very close to their true steady--state
shape. Investigation of the steady--state properties is therefore the
starting point also for the study of their time evolution. Moreover,
this suggests that a comparison of our steady--state shapes to shapes
obtained in field measurements is justified. In fact, the calculated
shapes agree nicely with recent measurements for barchan dunes
\cite{sauermann-etal:2000}. The situation is less clear for small
heaps, where mass losses can be of the order of the total flux and may
thus lead to significant differences between the steady--state and the
transient shapes under vanishing influx.

\begin{figure}[tb]
  \begin{center}  
  \psfrag{h in m}{$H$ [m]}
  \psfrag{delta in m}{$\delta x_{\tau, q}$ [m]}
  \psfrag{time}{time}
  \psfrag{xmaxtau}{$\delta x_\tau$}
  \psfrag{xmaxqxx}{$\delta x_q$}
  \includegraphics[width=\columnwidth]{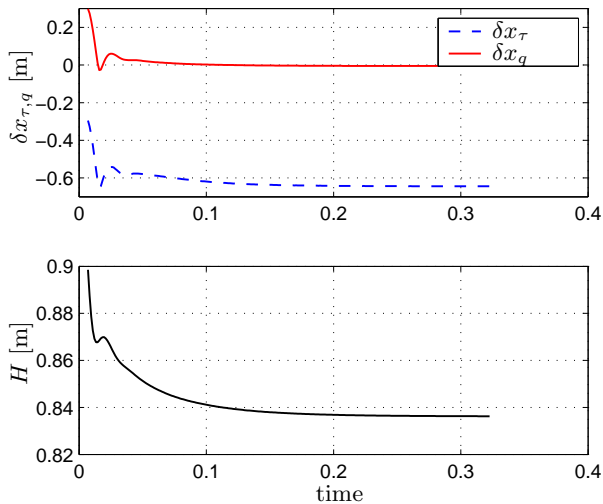} 
 \caption{The figure shows the transient evolution of various
  interesting length scales for a heap. \emph{Lower part:} Height of
  the heap. \emph{Upper part:} Distance of the locations of the
  maximum of the shear stress and of the maximum of the sand flux from
  the position of the top of the heap. In the steady state, the
  erosion/deposition vanishes at the crest.}  \label{fig:deltaheap}
\end{center}
\end{figure}

\begin{figure}[tb]
  \begin{center}  
  \psfrag{h in m}{$H$ [m]}
  \psfrag{delta in m}{$\delta x_{\tau, q}$ [m]}
  \psfrag{time}{time}
  \psfrag{xmaxtau}{$\delta x_\tau$}
  \psfrag{xmaxqxx}{$\delta x_q$}

  \includegraphics[width=\columnwidth]{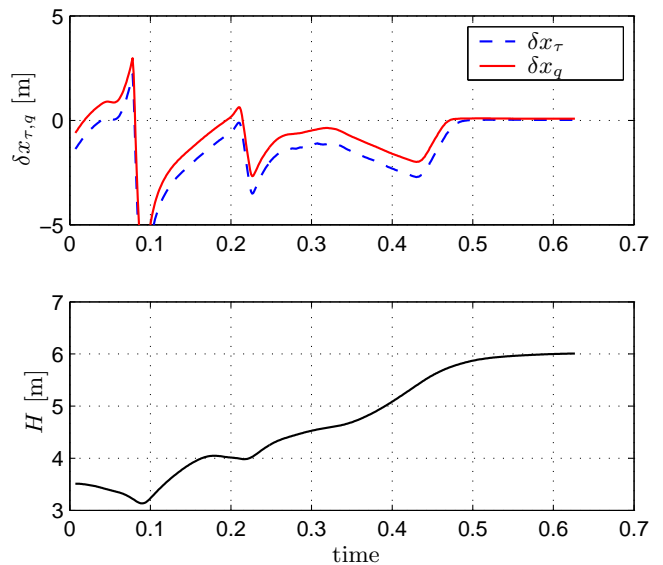}
  \caption{The figure shows the transient evolution of various
  interesting length scales for a dune that develops out of a smooth
  heap as in Fig.~\ref{fig:growth}. \emph{Lower part:} Height of the
  dune. \emph{Upper part:} Distance of the locations of the maximum of
  the shear stress and of the maximum sand flux from the position of
  the top of the crest. The lag between shear stress and sand flux
  vanishes, when the slip face reaches the crest.}
  \label{fig:deltadune}
\end{center}
\end{figure}

In the remainder of this section, we want to investigate more closely
the mechanism that drives the shape relaxation. As we have pointed
out, the positions of the maximum of the sand flux and of the maximum
of the profile must coincide in the steady state to make the
erosion/deposition vanish at the crest. We have shown that for small
heaps, this can be achieved by a fine--tuning of $\delta x_\tau$ to
about $\ell_s$. In contrast, for large dunes and strong winds, $\delta
x_\tau \gg \ell_s$, and the steady--state condition can only be met
with a singularity at the crest. This important difference is
exemplified by Figs.~\ref{fig:deltaheap} and \ref{fig:deltadune}. Both
figures show the evolution of the height and the displacements $\delta
x_\tau$ and $\delta x_q$ of the locations of the maximum of the shear
stress and of the maximum of the sand flux from the location of the
top of the sand profile, respectively. The distance between both
displacements is the lag of the flux with respect to the shear stress
due to the saturation transients, and is therefore closely related to
the saturation length $\ell_s$ for smooth surface profiles.  It
guarantees the proper vanishing of the erosion rate $q'$ at the top of
a steady--state heap where $\delta x_\tau$ is finite, but vanishes for
large steady--state dunes, where the slip face ends in a sharp brink
singularity at which the grains fall into the wake and are quickly
deposited.

 \noindent

\section{Summary and outview}\label{sec:sum}
In summary, we have shown that a simple minimal model for the
wind--driven sediment transport over a sand dune is capable of
explaining several important features of desert topographies. Among
them are the migration velocities of heaps and dunes, their shape
along the wind direction and the existence of a minimal dune size and
a maximum heap size.

As we have emphasized throughout this contribution and demonstrated by
the numerical solutions in the preceding section, the symmetry
breaking part of the shear stress exerted by turbulent air flow on an
obstacle, and local deviations of the sediment flux from its
equilibrium transport capacity (``saturation transients''), are the
essential ingredients in the modeling of aeolian sand dunes. It is
exactly the balance of these two relatively small effects that is
responsible for the relaxation of arbitrary initial conditions into a
characteristic dune or heap shape. Their neglect was responsible for
the failure of the naive zeroth order model discussed in
Section~\ref{sec:zero}. In hindsight we can say that it is not so much
the quantitative errors but the omission of this \emph{qualitatively}
important mechanism, what makes the zeroth order model an insufficient
description. In contrast, taking this balance properly into account,
makes the minimal model structurally stable against the neglect of
less significant quantitative details of the same order of magnitude.

\begin{figure}[tb]
  \psfrag{L}{$L$ [m]}
  \psfrag{ustar}{$u_*/u_{*t}$}
  \begin{center}
  \includegraphics[width=\columnwidth]{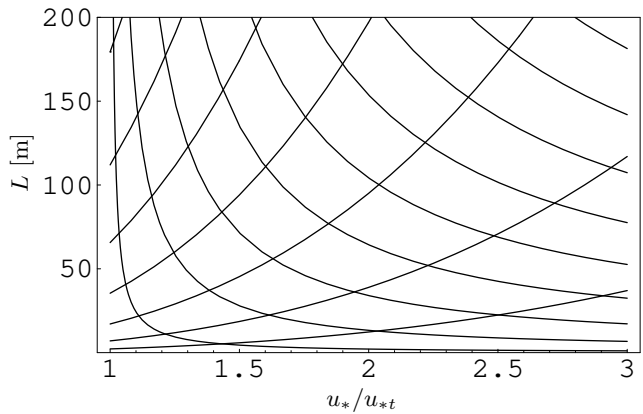}
  \caption{Qualitative shape diagram that could be useful
  in the analysis and comparison of field studies. The migration
  velocity is constant along rising lines, whereas falling lines
  indicate invariant dune shape.} \label{fig:shapediagram}
\end{center}
\end{figure}

This direction was recently pursued further in an effort to calculate
analytically certain steady state shapes of dunes and heaps by
``linearizing'' the minimal model \cite{andreotti-claudin:2002}. One
may as well wish to proceed also into the opposite direction. After
the basic mechanism is understood, more elaborate dune models can be
constructed by putting some of the neglected details back into the
description. Detailed parametric studies of such a refined model for a
certain dune type and comparison to field data would be very useful to
test some of the less generic predictions of the underlying sand
transport model \cite{sauermann-kroy-herrmann:2001}, such as the shear
stress dependence of the saturation length $\ell_s$
(Fig.~\ref{fig:ls}). This is important, since, as we have shown, the
variable parameter $\ell_s$ sets the characteristic length scale with
respect to which dunes and heaps can be said to be large or
small. Phenomenological knowledge about $\ell_s$ is still very
limited. More detailed studies could further be helpful to map out
quantitative shape diagrams, of the type sketched qualitatively in
Fig.~\ref{fig:shapediagram}. These diagrams could be useful not only
for the validation of the model, but also for the comparison of field
data from different places with different prevailing wind and sand
conditions. The migration velocity is constant along the rising lines
in Fig.~\ref{fig:shapediagram}, which were obtained from
Eq.(\ref{eq:v}) using $q\approx q_s$ together with
Eq.(\ref{eq:ls}). They allow for example a comparison of the migration
velocities of dunes of different sizes that are exposed to (on
average) identical winds. Or one may infer the average wind speed from
measurements of sizes and migration velocities in a dune field. The
falling lines in Fig.~\ref{fig:shapediagram} are lines of constant
shape, assuming that the latter is determined by $\ell_s/L$. They may
thus be used for correlating wind speeds with dune shapes. In general
(in particular for the full $3d$ problem), such shape diagrams will be
more complex since the influx is an additional important variable that
we have neglected here, as it vanishes for $2d$ dunes in the steady
state.

Moreover, as we pointed out, there are still many consequences of the
present model that await a systematic investigation.  And a major
future task is finally the generalization of the present discussion to
the $3d$ case. A promising route could be the construction of an
effectively sliced model that allows one to use the proposed model for
the separation bubble and to keep the time--limiting calculation (the
integration of the flux equation) effectively one--dimensional. The
smaller transverse currents could be inferred from the sliced
solution. A generalization of the flux equation to the $2d$ surface of
a $3d$ dune is also feasible \cite{sauermann:phd}.  A more ambitious
task will eventually be the simulation of dune fields.  Whereas the
existence of a minimum dune size could be obtained by an analysis of
the shape stability alone, the question of a possible existence of a
characteristic or maximum dune size in a dune field, depends on the
mass balance of a dune in the complicated environment provided by the
other dunes, and is much more difficult to answer
\cite{lima-etal:2002}.

\begin{acknowledgments}
We gratefully acknowledge financial support by the Deutsche
Forschungsgemeinschaft through contract No HE 2732/1-1, and helpful
discussions with Ken Andersen and Philippe Claudin.
\end{acknowledgments}


\begin{thebibliography}{40}
\bibitem[{spe()}]{speed}
\bibinfo{note}{This has to be distinguished from the statement that the
  migration velocity of a dune with slip face is proportional to the flux over
  the brink divided by the height, which is (trivially) true if losses over the
  slip face are negligible, but useless unless the flux over the brink is
  specified.}
\bibitem[{bad()}]{badbubble}
\bibinfo{note}{An inapropriate boundary condition appears to be the main reason
  for the failure of the original attempts to model flow separation in this way
  \cite{zeman-jensen:88}.}
\end{thebibliography}

\end{document}